\documentclass[11pt,preprint]{elsarticle}
\usepackage[T1]{fontenc}
\usepackage[latin9]{inputenc}
\usepackage[a4paper]{geometry}
\usepackage{color}
\usepackage{mathtools}
\usepackage{enumitem}
\usepackage{amsmath}
\usepackage{amsthm}
\usepackage{amssymb}
\usepackage{wasysym}
\usepackage[unicode=true,
 bookmarks=false,
 breaklinks=true,pdfborder={0 0 0},pdfborderstyle={},backref=false,colorlinks=true]
 {hyperref}
\hypersetup{pdftitle={New boundary monodromy matrices for classical sigma models},
 pdfauthor={Tamás Gombor }}

\makeatletter

\providecommand{\tabularnewline}{\\}

\theoremstyle{plain}
\newtheorem*{thm*}{\protect\theoremname}

\newcommand{\gn}{\mathfrak{g}}
\newcommand{\hn}{\mathfrak{h}}
\newcommand{\fn}{\mathfrak{f}}
\newcommand{\ue}{\mathfrak{u}(1)}
\newcommand{\su}[1]{\mathfrak{su}(#1)}
\newcommand{\so}[1]{\mathfrak{so}(#1)}
\newcommand{\spp}[1]{\mathfrak{sp}(#1)}

\newcommand{\SU}[1]{\mathrm{SU}(#1)}
\newcommand{\SO}[1]{\mathrm{SO}(#1)}
\newcommand{\On}[1]{\mathrm{O}(#1)}
\newcommand{\Sp}[1]{\mathrm{Sp}(#1)}

\newcommand{\Tr}{\mathrm{Tr}}
\newcommand{\dd}{\mathrm{d}}
\newcommand{\Jc}{\mathrm{J}}
\newcommand{\Ic}{\hat{\mathrm{J}}}
\newcommand{\Lc}{\mathrm{L}}

\newcommand{\di}{\mathrm{d}}

\newcommand{\n}{\mathbf{n}}
\newcommand{\dal}{\dot{\alpha}}
\newcommand{\dbe}{\dot{\beta}}
\newcommand{\bsig}{\bar{\sigma}}

\newcommand{\nt}{\tilde{\mathbf{n}}}
\newcommand{\nh}{\hat{\mathbf{n}}}

\newcommand{\Pexp}{\mathcal{P}\overleftarrow{\exp}}

\makeatother

\providecommand{\theoremname}{Theorem}

\begin{document}

\begin{frontmatter}{}

\title{New boundary monodromy matrices for classical sigma models}

\author{Tamás Gombor}

\ead{gombor.tamas@wigner.mta.hu}

\address{Lendület Holographic QFT Group, Wigner Research Centre for Physics,
Konkoly-Thege Miklós u. 29-33, 1121 Budapest , Hungary}
\begin{abstract}
The 2d principal models without boundaries have $G\times G$ symmetry.
The already known integrable boundaries have either $H\times H$ or
$G_{D}$ symmetries, where $H$ is such a subgroup of $G$ for which
$G/H$ is a symmetric space while $G_{D}$ is the diagonal subgroup
of $G\times G$. These boundary conditions have a common feature:
they do not contain free parameters. We have found new integrable
boundary conditions for which the remaining symmetry groups are either
$G\times H$ or $H\times G$ and they contain one free parameter.
The related boundary monodromy matrices are also described. 
\end{abstract}

\end{frontmatter}{}

\tableofcontents{}

\section{Introduction}

In this paper we investigate $1+1$ dimensional $\On{N}$ sigma and
principal chiral models (PCMs). These are integrable at the quantum
level i.e. infinite many conserved charges survive the quantization
\citep{Polyakov:1977vm,Goldschmidt:1980wq}. The scattering matrices
(S-matrices) are factorized and they can be constructed from the two
particle S-matrices which satisfy the Yang-Baxter equation (YBE).
Thus, integrable theories at infinite volume can be defined by the
solutions of the YBE. For example, it has been verified that the minimum
solution of the $\On{N}$ symmetric YBE is the S-matrix of the $\On{N}$
sigma model \citep{Zamolodchikov:1978xm}. 

In this paper we are interested in boundary conditions for these systems.
There are three interesting type of boundary conditions which are:
\begin{enumerate}[label=\Roman*]
\item Classically conformal - which means that the boundary condition does
not break the classical conformal symmetry, which guaranties infinitely
many conserved charges 
\item Boundary conditions with zero curvature representation which means
that there exists a $\kappa$-matrix (or classical reflection matrix)
from which double row monodromy matrices can be constructed 
\item Quantum integrable, which means that there exist a higher spin conserved
charge even on the half line.
\end{enumerate}
The basic examples of boundary conditions of $\On{N}$ sigma models
are:
\begin{enumerate}
\item Restricted boundary conditions when we restrict the field to a lower
dimensional sphere
\begin{enumerate}
\item with arbitrary radius\label{enu:SigResA}
\item with maximal radius\label{enu:SigResM}
\end{enumerate}
\item Using boundary Lagrangian $L_{b}=\n^{T}M\dot{\n}$ with $M\in\mathfrak{so}(N)$
(See notations in Section \ref{sec:-sigma-model})
\begin{enumerate}
\item where $M$ is arbitrary \label{enu:SigLagA}
\item where $M^{2}=c1$ \label{enu:SigLagM}
\item where $M^{2}=\mathrm{diag}(c,c,0,\dots,0)$. \label{enu:SigLagN}.
\end{enumerate}
\end{enumerate}
The basic examples of boundary conditions of PCM of group $G$ are
as follows:
\begin{enumerate}[label=\roman*]
\item Restricted boundary condition when we restrict the field to a subgroup
$H$.
\begin{enumerate}
\item where $H$ is arbitrary\label{enu:PCMResA}
\item where $G/H$ is a symmetric space.\label{enu:PCMResM}
\end{enumerate}
\item Using boundary Lagrangian $L_{b}=\mathrm{Tr}\left(MJ_{0}\right)$
with $M\in\mathfrak{g}$ (See notations in Section \ref{sec:Principal-Chiral-Models})
\begin{enumerate}
\item where $M$ is arbitrary \label{enu:PCMLagA}
\item where the $G/H$ is a symmetric space for $H\coloneqq\left\{ h\in G|hMh^{-1}=M\right\} $.
\label{enu:PCMLagM}.
\end{enumerate}
\end{enumerate}
These boundary conditions were investigated in \citep{Corrigan:1996nt,MacKay:1999ur,MacKay:2001bh,Moriconi:2001xz,Mann:2006rh,Aniceto:2017jor}
and was shown that all of them are conformal. What can we say about
the quantum integrability of these boundary conditions? In some of
these cases, one can also use the Goldschmidt-Witten argument \citep{MacKay:1999ur,Moriconi:2001xz}
which is a sufficient condition for quantum integrability. With this
argument it can be shown that boundary conditions \ref{enu:SigResM}
and \ref{enu:PCMResM} are integrable at the quantum level.

There is also a necessary condition for quantum integrability which
comes from the boundary bootstrap. As we know, quantum integrable
theories with boundary can be defined with the bulk S-matrix and the
boundary scattering matrix (or reflection matrix, R-matrix). Reflection
matrices are solutions of the boundary Yang-Baxter equation (bYBE).
They are classified for the $\On{N}$ sigma model \citep{Moriconi:2001xz,Arnaudon:2003gj}.
There are two classes which have symmetries either $\On{k}\times\On{N-k}$
or $\mathrm{U}(n)$ if $N=2n$. There is a free parameter in the reflection
matrix when the remaining symmetries are $\On{2}\times\On{N-2}$ and
$\mathrm{U}(n)$. Thus we can infer that if the center of the residual
symmetry algebra is $\ue$ then the reflection matrix contains a free
parameter \citep{Gombor:2017qsy}.

We can also classify the residual symmetries of PCMs. The bulk theory
has $G_{L}\times G_{R}$ symmetry and the particles transform with
respect to some representations of this symmetry. If the reflection
matrix has a factorized form ($R=R_{L}\otimes R_{R}$), then the bYBE
can be separated into equation for left and right reflection matrices.
Thus, in principle, arbitrarily combined solutions $R_{L}$ and $R_{R}$
can be used to construct the full reflection matrix $R$. This implies
that the remaining left and right symmetries can be different. 

From the classification of the quantum reflection matrices \citep{Moriconi:2001xz,MacKay:2001bh,Arnaudon:2003gj,Gombor:2017qsy,Gombor:2019bun}
we can extract the possible residual symmetries therefore we can conclude
that \ref{enu:SigResA}, \ref{enu:SigLagA}, \ref{enu:PCMResA} and
\ref{enu:PCMLagA} can not be quantum integrable because their residual
symmetries are different.

The zero curvature description is also known for some boundary conditions
\citep{Corrigan:1996nt,Mann:2006rh}. Their classical reflection matrices
are constant matrices without any parameters. 
\begin{table}
\begin{centering}
\begin{tabular}{|c|c|c|c|}
\hline 
 & I & II & III\tabularnewline
\hline 
\hline 
\ref{enu:SigResA} & $\checked$ & $\times$ & $\times$\tabularnewline
\hline 
\ref{enu:SigResM} & $\checked$ & $\checked$ & $\checked$\tabularnewline
\hline 
\ref{enu:SigLagA} & $\checked$ & $?$ & $\times$\tabularnewline
\hline 
\ref{enu:SigLagM} & $\checked$ & $?$ & $?$\tabularnewline
\hline 
\ref{enu:SigLagN} & $\checked$ & $?$ & $?$\tabularnewline
\hline 
\ref{enu:PCMResA} & $\checked$ & $\times$ & $\times$\tabularnewline
\hline 
\ref{enu:PCMResM} & $\checked$ & $\checked$ & $\checked$\tabularnewline
\hline 
\ref{enu:PCMLagA} & $\checked$ & $?$ & $\times$\tabularnewline
\hline 
\ref{enu:PCMLagM} & $\checked$ & $?$ & $?$\tabularnewline
\hline 
\end{tabular}
\par\end{centering}
\caption{Properties of boundary conditions}
\label{tab:tab1}
\end{table}

The state of the art about boundary conditions and their integrability
can be summarized in Table \ref{tab:tab1}. With question marks we
indicated the open questions. For example, \ref{enu:SigResM} is quantum
integrable (Goldschmidt-Witten argument) and it has $\On{k}\times\On{N-k}$
symmetry so it can be matched to the reflection matrix (coming from
the bootstrap) with the same symmetry. Contrary, we have a $\mathrm{U}(N/2)$
symmetric reflection matrix with a free parameter and one can ask
which boundary condition belongs to it. The boundary condition \ref{enu:SigLagM}
is a natural candidate because it has a free parameter and the same
symmetry. Indeed, in this paper we show that it has a zero curvature
representation which may indicate the quantum integrability in view
of the fact that a restricted boundary condition preserved the integrability
at the quantum level if and only if there exists a zero curvature
representation (see the table above).

In the PCM the remaining symmetries for the known classical integrable
boundary conditions are $H_{L}\times H_{R}$ where $H_{L}\cong H_{R}$
which means $R_{L}\cong R_{R}$ (or the residual symmetry is $G_{D}$
which is the diagonal subgroup of $G_{L}\times G_{R}$ but in this
case the reflection matrix is not factorized) \citep{MacKay:1999ur,MacKay:2001bh}.
This paper also provides a zero curvature representation for boundary
condition \ref{enu:PCMLagM} where only the left or the right symmetries
are broken therefore these can be candidates for reflection matrices
where $R_{L}\not\cong R_{R}$.

We also derive that the traces of these new monodromy matrices Poisson
commute therefore there are infinitely many conserved charges in involution.
This Poisson algebra of the one and double row monodromy matrices
are consistent if the $r$-matrix and classical reflection matrix
($\kappa$-matrix) satisfy the classical Yang-Baxter (cYBE) and the
classical boundary Yang-Baxter equations (cbYBE). In \citep{Corrigan:1996nt}
and \citep{Mann:2006rh} the Poisson algebra was investigated for
non-ultralocal theories with constant $\kappa$-matrix. In \citep{Avan:2018pua}
this was done for ultralocal theories with dynamical $\kappa$-matrix
when the Poisson bracket of the $\kappa$-matrix and the Lax-connection
vanished. In this paper we derive the Poisson algebra of non-ultralocal
theories with $\kappa$-matrix whose Poisson-bracket with the Lax-connection
does not vanish. However, the possible solutions of this equation
have only been examined in a few cases. In this paper we classify
the solutions of the field independent cbYBE and check that the new
field dependent $\kappa$-matrix is satisfies the cbYBE for $\mathrm{O}(N)$
sigma models.

The paper is structured as follows. In the next section, we start
with the Lax formalism of the PCMs where we construct classical reflection
matrices and use them to build double row transfer matrices. The conservation
of these matrices (which is equivalent to the existence of infinite
many conserved charges) provides the boundary conditions of the theories
which belong to these boundary Lax representations. Using these results,
we derive new double row monodromy matrices for the $\On{2n}$ sigma
models and the corresponding boundary conditions will be determined
too. In Section \ref{sec:Poisson-algebra-of} we derive the Poisson
algebra of the double row monodromy matrices and the cbYBE which is
satisfied for the new $\kappa$-matrices.

\section{Principal Chiral Models on the half line\label{sec:Principal-Chiral-Models}}

In this section the new boundary monodromy matrix will be introduced.
In the first subsection we will overview the Lax formalism of PCMs.
After that the new reflection matrix and the related boundary condition
will be derived. Finally we will show the corresponding Lagrangian
descriptions and the unbroken symmetries of these models.

\subsection{Lax formalism for PCMs\label{subsec:Lax-formalism-for}}

Let $\gn$ be a semi-simple Lie algebra and $G=\exp(\gn)$. We use
only matrix Lie-algebra and we work in the defining representation.
The field variable is a map $g:\Sigma\rightarrow G$ where the space-time
$\Sigma=\mathbb{R}\times(-\infty,0]$ is parameterized with $(x^{0},x^{1})=(t,x)$.
We can define two currents $\Jc^{R}=g^{-1}\dd g$ and $\Jc^{L}=g\dd g^{-1}$
where $\Jc^{L/R}=J_{0}^{L/R}\dd x^{0}+J_{1}^{L/R}\dd x^{1}\left(=J_{t}^{L/R}\dd t+J_{x}^{L/R}\dd x\right)$\footnote{The ordinary letters denote forms and the italic letters denote the
local coordinate functions of these. }. These two currents satisfy the flatness condition (by definition):
\[
\dd\Jc^{L/R}+\Jc^{L/R}\wedge\Jc^{L/R}=0
\]
The bulk equation of motion (E.O.M) is 
\[
\dd*\Jc^{L/R}=0.
\]
The E.O.M and the flatness condition is equivalent to the flatness
condition of the Lax connection: 
\begin{equation}
\dd\Lc^{L/R}(\lambda)+\Lc^{L/R}(\lambda)\wedge\Lc^{L/R}(\lambda)=0\label{eq:zerocurv}
\end{equation}
where 
\[
\Lc^{L/R}(\lambda)=\frac{1}{1-\lambda^{2}}\Jc^{L/R}+\frac{\lambda}{1-\lambda^{2}}*\Jc^{L/R}.
\]
We will also use the following notations
\begin{align*}
\mathcal{M}^{L/R}(\lambda) & =L_{0}^{L/R}(\lambda) & \mathcal{L}^{L/R}(\lambda) & =L_{1}^{L/R}(\lambda)
\end{align*}
Using these, the zero curvature condition can be written as
\[
\partial_{t}\mathcal{L}(\lambda)-\partial_{x}\mathcal{M}(\lambda)+\left[\mathcal{M}(\lambda),\mathcal{L}(\lambda)\right]=0.
\]

The usefulness of the Lax connection lies in the fact that one can
generate from it an infinite family of conserved charges. At first
we define the one row monodromy matrix 
\begin{equation}
T_{L/R}(\lambda)=\Pexp\left(-\int_{-\infty}^{0}\mathcal{L}^{L/R}(\lambda)\dd x\right).\label{eq:globT}
\end{equation}
These monodromy matrices have an inversion property 
\begin{equation}
T_{R}(\lambda)=g^{-1}(0)T_{L}(1/\lambda)g(-\infty).\label{eq:invprop}
\end{equation}
The monodromy matrix in the boundary case takes a double row type
form 
\begin{equation}
\Omega_{L/R}(\lambda)=T_{L/R}(-\lambda)^{-1}\kappa_{L/R}(\lambda)T_{L/R}(\lambda),\label{eq:globO}
\end{equation}
where the $\kappa_{L}(\lambda),\kappa_{R}(\lambda)\in G$ are the
reflection matrices which will be specified later. In the following
we use the right currents therefore we introduce the following notation
$\Jc(\lambda)=\Jc^{R}(\lambda)$, $\Lc(\lambda)=\Lc^{R}(\lambda)$,
$T(\lambda)=T_{R}(\lambda)$, $\Omega(\lambda)=\Omega_{R}(\lambda)$,
$\kappa(\lambda)=\kappa_{R}(\lambda)$, $\mathcal{M}(\lambda)=\mathcal{M}^{R}(\lambda)$
and $\mathcal{L}(\lambda)=\mathcal{L}^{R}(\lambda)$

The existence of infinitely many conserved quantities requires that
the time derivative of the monodromy matrix has to vanish $\dot{\Omega}(\lambda)=0$,
which is equivalent to: 
\begin{equation}
\kappa(\lambda)\mathcal{M}(\lambda)\Big|_{x=0}-\mathcal{M}(-\lambda)\Big|_{x=0}\kappa(\lambda)=\dot{\kappa}(\lambda),\label{eq1}
\end{equation}
where we assumed that the currents vanish at $-\infty$. This is the
\textit{boundary flatness condition}.

This equation can be translated to boundary conditions for the $\Jc^{R}$
current. The consistency of the theory requires that the number of
boundary conditions have to be equal to $\mathrm{dim}(\gn)$. Based
on these, we call $\kappa(\lambda)$ a consistent solution of \eqref{eq1}
if it leads to exactly $\mathrm{dim}(\gn)$ boundary conditions.

The consistency of the definitions of double row monodromy matrices
$\Omega_{L}$ and $\Omega_{R}$ (the boundary flatness condition implies
the same boundary conditions with $\Omega_{L}$ and $\Omega_{R}$)
implies that
\begin{equation}
\kappa_{R}(\lambda)=g^{-1}(0)\kappa_{L}(1/\lambda)g(0).\label{eq:invKappa}
\end{equation}
Using this equation, the double row monodromy matrices also have an
inversion property:
\begin{equation}
\Omega_{R}(\lambda)=g^{-1}(-\infty)\Omega_{L}(1/\lambda)g(-\infty).\label{eq:invOmega}
\end{equation}

Hereinafter, we look for consistent solutions for the equation \eqref{eq1}.
The most obvious ansatz for the reflection matrix is $\kappa(\lambda)=U$
where $U\in G$ is a constant matrix. Using this ansatz, the equation
$\eqref{eq1}$ is equivalent to the following two equations: 
\begin{align*}
J_{0} & =UJ_{0}U^{-1},\\
-J_{1} & =UJ_{1}U^{-1}.
\end{align*}
Clearly, $J_{0}$ and $J_{1}$ are elements of the eigenspaces of
the linear transformation $\mathrm{Ad}_{U}:\gn\rightarrow\gn$ with
$+1$ and $-1$ eigenvalues. These are equivalent to $\dim(\gn)$
boundary conditions if and only if $U^{2}$ is proportional to 1.
Thus, there is a $\mathbb{Z}_{2}$ graded decomposition $\gn=\hn\oplus\fn$
where $\hn$ and $\fn$ are the $+1$ and $-1$ eigenspaces of the
$\mathrm{Ad}_{U}$ automorphism of $\gn$. Therefore the boundary
conditions imply $J_{0}\in\hn$ and $J_{1}\in\fn$. These are well
known integrable boundary conditions \citep{MacKay:2001bh}. In the
next subsection, we will try to find new consistent solutions with
non-trivial spectral parameter dependency.

Before that, we note that there is another possibility for the definition
of the double row monodromy matrix, namely: 
\[
\Omega(\lambda)=T_{L}(-\lambda)^{-1}UT_{R}(\lambda),
\]
This leads to the following boundary conditions 
\begin{align}
J_{0}^{L} & =UJ_{0}^{R}U^{-1},\label{bcnonch}\\
-J_{1}^{L} & =UJ_{1}^{R}U^{-1}.\label{eq:bcnonch}
\end{align}
Let us calculate the number of boundary conditions. For this, let
us use the relation between the left and right currents.
\begin{align*}
-gJ_{0}^{R}g^{-1} & =UJ_{0}^{R}U^{-1}, & \Rightarrow &  & -J_{0}^{R} & =\left(U^{-1}g\right)^{-1}J_{0}^{R}U^{-1}g,\\
+gJ_{1}^{R}g^{-1} & =UJ_{1}^{R}U^{-1}, & \Rightarrow &  & +J_{1}^{R} & =\left(U^{-1}g\right)^{-1}J_{1}^{R}U^{-1}g.
\end{align*}
We saw previously that this type of boundary condition is consistent
if the operator $\mathrm{Ad}_{U^{-1}g}$ is an involution on $\mathfrak{g}$
which is equivalent to 
\begin{equation}
U^{-1}gU^{-1}g=e.\label{eq:adj}
\end{equation}
Clearly this restricted boundary condition is invariant under the
transformation $g\to Ug_{0}^{-1}U^{-1}gg_{0}$ therefore it has the
diagonal symmetry $G_{D}$.

Finally, let us note that there is an other representation of this
boundary condition. Using the inversion property \eqref{eq:invprop}
we can obtain an equivalent double row monodromy matrix:
\[
\Omega(\lambda)=T_{R}(-1/\lambda)^{-1}\left(g(0)^{-1}U\right)T_{R}(\lambda),
\]
The conservation of this double row monodromy matrix requires that
the following boundary flatness condition has to vanish. 
\[
g^{-1}U\left(J_{0}^{R}-\lambda J_{1}^{R}\right)+\left(\lambda^{2}J_{0}^{R}+\lambda J_{1}^{R}\right)g^{-1}U=(1-\lambda^{2})\partial_{0}\left(g^{-1}U\right)=(\lambda^{2}-1)J_{0}^{R}g^{-1}U
\]
Multiplying this by $g$ from the right, we obtain
\[
U\left(J_{0}^{R}-\lambda J_{1}^{R}\right)+\lambda gJ_{1}^{R}g^{-1}U=-gJ_{0}^{R}g^{-1}U
\]
which leads to the equations \eqref{bcnonch} and \eqref{eq:bcnonch}.

\subsection{Spectral parameter dependent $\kappa$-matrices\label{subsec:Spectral-parameter-dependent}}

In the previous subsection we summarized the spectral parameter independent
$\kappa$-matrices. In this subsection, we try to find new \textit{spectral
parameter dependent} $\kappa$s. 

\subsubsection{Solution of the boundary flatness equation}

Let us use the following ansatz: 
\begin{equation}
\kappa(\lambda)=k(\lambda)(1+\lambda M+\lambda^{2}N),\label{eq0}
\end{equation}
where $k(z)$ is a scalar and $M\in\gn$. Using this ansatz the equation
$\eqref{eq1}$ takes the following form: 
\[
\left(1+\lambda M+\lambda^{2}N\right)(J_{0}-\lambda J_{1})-(J_{0}+\lambda J_{1})\left(1+\lambda M+\lambda^{2}N\right)=0.
\]
Which leads to the following system of equations: 
\begin{align}
\lambda^{1}: &  & [M,J_{0}]-2J_{1} & =0\label{eqq1}\\
\lambda^{2}: &  & [N,J_{0}]-\left[M,J_{1}\right]_{+} & =0\label{eqq2}\\
\lambda^{3}: &  & \left[N,J_{1}\right]_{+} & =0\label{eqq3}
\end{align}
where $\left[,\right]_{+}$ is the anti-commutator i.e. $\left[X,Y\right]_{+}=XY+YX$.
Since equation \eqref{eqq1} provides already $\dim(\gn)$ boundary
conditions, the consistency requires that the equations \eqref{eqq2}
and \eqref{eqq3} should follow from \eqref{eqq1}. In the following,
we look for constraints on $M$ and $N$ which ensure this.

Taking the anti-commutator of equation \eqref{eqq1} with $M$ gives
\[
\left[M,J_{1}\right]_{+}=\frac{1}{2}\left[M,\left[M,J_{0}\right]\right]_{+}=\frac{1}{2}\left[M^{2},J_{0}\right].
\]
The r.h.s is equal to $[N,J_{0}]$ if 
\begin{equation}
N-\frac{1}{2}M^{2}=c1,\label{cond1}
\end{equation}
where $c$ is a constant. From this we can see that $M$ commutes
with $N$. Using this and the equation \eqref{eqq3} we can obtain:
\[
\left[N,\left[M,J_{1}\right]_{+}\right]_{+}=0.
\]
Therefore, by taking the anti-commutator of equation \eqref{eqq2}
with $N$, we get
\[
\left[N^{2},J_{0}\right]=0.
\]
Since $J_{0}$ spans the whole defining representation of $\mathfrak{g}$
therefore $N^{2}$ has to be proportional to $1$ so the automorphism
$\mathrm{Ad}_{N}$ has $+1$ and $-1$ eigenvalues and we denote the
corresponding eigenspaces by $\hn$ and $\fn$. Therefore $N$ defines
a $\mathbb{Z}_{2}$ graded decomposition $\gn=\hn\oplus\fn$.

Equation \eqref{eqq3} means that $J_{1}\in\fn$ i.e $\Pi_{\hn}(J_{1})=0$
where $\Pi_{\hn}$ is the projection operator of $\hn$ subspace.
Putting this into \eqref{eqq2}: 
\[
\Pi_{\hn}\left(J_{1}\right)=\frac{1}{2}\Pi_{\hn}\left([M,J_{0}]\right)=\frac{1}{2}\left[M,\Pi_{\hn}(J_{0})\right]
\]
where we used that $[M,N]=0$ which implies $M\in\hn$. We can see
from the last equation that equation \eqref{eqq3} follows from \eqref{eqq1}
if $M$ commutes with $\hn$.

Summarizing, consistency of the solutions requires the following conditions
\begin{align}
2N-M^{2} & \sim1 & \text{and} &  & N^{2} & \sim1.\label{condMN}
\end{align}
These implies that $\mathrm{Ad}_{N}$ generates a $\mathbb{Z}_{2}$
graded decomposition and $M$ is an element of $\hn$ and also commutes
with $\hn$. Therefore $\hn$ has a non-trivial center which is generated
by $M$. It follows that every $\mathbb{Z}_{2}$ graded decomposition
where $\hn$s are not semi-simple belong to these type of reflection
matrices and boundary conditions.

There are two classes of these $\kappa$ matrices. The first is $N\neq0$.
The second case is $N=0$, which implies that $M^{2}\sim1$. In this
case $M$ defines the $\mathbb{Z}_{2}$ graded decomposition. The
projection operators to the $\mathfrak{h}$ and $\mathfrak{f}$ are:
\begin{align*}
\Pi_{\mathfrak{h}}(X) & =\frac{1}{2}\left(X+UXU^{-1}\right),\\
\Pi_{\mathfrak{f}}(X) & =\frac{1}{2}\left(X-UXU^{-1}\right),
\end{align*}
where $U=N$ when $N\neq0$ otherwise $U=M$. The classification of
these $\kappa$-matrices for classical Lie-algebras are shown in the
following.

\subsubsection{Examples}

We saw that the integrable boundary conditions described above belongs
to a $(\mathfrak{g},\mathfrak{h})$ symmetric pair for which $G/H$
is a symmetric spaces ($G=\exp(\mathfrak{g})$, $H=\exp(\mathfrak{h})$).
The symmetric spaces are classified \citep{helgason1979differential}.
The spectral parameter dependent solutions belongs to not semi-simple
$\mathfrak{h}$ therefore there are three types of spectral parameter
dependent $\kappa$-matrices.
\begin{enumerate}
\item $\gn=\su{n}$ and $\hn=\ue\oplus\su{m}\oplus\su{n-m}$. The $\ue\subset\hn$
sub-algebra is generated by the matrix $M$ and condition \eqref{condMN}
leads to the following $N$: 
\begin{align*}
M & =i\frac{2a}{k-m}\begin{pmatrix}-k1_{m\times m} & 0_{m\times k}\\
0_{k\times m} & m1_{k\times k}
\end{pmatrix}, & N & =a^{2}\frac{n}{k-m}\begin{pmatrix}-1_{m\times m} & 0_{m\times k}\\
0_{k\times m} & 1_{k\times k}
\end{pmatrix},
\end{align*}
where $k=n-m$. One can choose a function $k(\lambda)$ for which
$\kappa(z)\in\mathrm{U}(n)$ when $z\in\mathbb{R}$: 
\[
\kappa_{1}(\lambda|a)=\begin{pmatrix}\frac{1+ia\lambda}{1-ia\lambda}1_{m\times m} & 0_{m\times k}\\
0_{k\times m} & 1_{k\times k}
\end{pmatrix}.
\]
\item $\gn=\so{n}$ and $\hn=\so{2}\oplus\so{n-2}$. The $M$, $N$ and
the $\kappa(\lambda)\in\SO{n}$ can be written as: 
\begin{align*}
M & =2a\begin{pmatrix}0 & -1 & 0 & 0 & \cdots\\
1 & 0 & 0 & 0 & \cdots\\
0 & 0 & 0 & 0 & \cdots\\
0 & 0 & 0 & 0 & \cdots\\
\vdots & \vdots & \vdots & \vdots & \ddots
\end{pmatrix}, & N & =a^{2}\begin{pmatrix}-1 & 0 & 0 & 0 & \cdots\\
0 & -1 & 0 & 0 & \cdots\\
0 & 0 & 1 & 0 & \cdots\\
0 & 0 & 0 & 1 & \cdots\\
\vdots & \vdots & \vdots & \vdots & \ddots
\end{pmatrix},
\end{align*}
\[
\kappa_{2}(\lambda|a)=\begin{pmatrix}A(\lambda|a) & -B(\lambda|a) & 0 & 0 & \cdots\\
B(\lambda|a) & A(\lambda|a) & 0 & 0 & \cdots\\
0 & 0 & 1 & 0 & \cdots\\
0 & 0 & 0 & 1 & \cdots\\
\vdots & \vdots & \vdots & \vdots & \ddots
\end{pmatrix},
\]
where 
\begin{align*}
A(\lambda|a) & =\frac{1-\lambda^{2}a^{2}}{1+\lambda^{2}a^{2}},\\
B(\lambda|a) & =\frac{2\lambda a}{1+\lambda^{2}a^{2}}.
\end{align*}
\item $\gn=\so{2n}$ or $\gn=\spp{n}$ and $\hn=\ue\oplus\su{n}$ For this
case 
\[
M=a\begin{pmatrix}0_{n\times n} & -1_{n\times n}\\
1_{n\times n} & 0_{n\times n}
\end{pmatrix}.
\]
Since $M^{2}=-a^{2}1$ then $N=0$. The $\kappa$-matrix is the following:
\[
\kappa_{3}(\lambda|a)=\frac{1}{\sqrt{1+\lambda^{2}a^{2}}}\begin{pmatrix}1_{n\times n} & -\lambda a1_{n\times n}\\
\lambda a1_{n\times n} & 1_{n\times n}
\end{pmatrix}.
\]
We can check that $\kappa_{3}(\lambda)\in\SO{2n}$ and $\kappa_{3}(\lambda)\in\Sp{n}$
too.
\end{enumerate}
These matrices are the classical counterparts of the $\hn=\ue\oplus\su{m}\oplus\su{n-m}$,
$\hn=\ue\oplus\su{n}$ and $\hn=\so{2}\oplus\so{n-2}$ symmetric solutions
of the quantum boundary Yang-Baxter equation \citep{Moriconi:2001xz}\citep{MacKay:2001bh}\citep{Aniceto:2017jor}.
The quantum reflection matrices are 
\begin{align*}
R_{1}(\theta|c) & =\nu_{1}(\theta|c)\begin{pmatrix}\frac{c-\theta}{c+\theta}1_{m\times m} & 0_{m\times k}\\
0_{k\times m} & 1_{k\times k}
\end{pmatrix},\\
R_{2}(\theta|c) & =\nu_{2}(\theta|c)\begin{pmatrix}\tilde{A}(\theta|c) & -\tilde{B}(\theta|c) & 0 & 0 & \cdots\\
\tilde{B}(\theta|c) & \tilde{A}(\theta|c) & 0 & 0 & \cdots\\
0 & 0 & 1 & 0 & \cdots\\
0 & 0 & 0 & 1 & \cdots\\
\vdots & \vdots & \vdots & \vdots & \ddots
\end{pmatrix},\\
R_{3}(\theta|c) & =\nu_{3}(\theta|c)\begin{pmatrix}c1_{n\times n} & -i\theta1_{n\times n}\\
i\theta1_{n\times n} & c1_{n\times n}
\end{pmatrix},
\end{align*}
where $\nu_{i}(\theta)$ are some dressing phases and 
\begin{align*}
\tilde{A}(\theta|c) & =\frac{1}{2}\left(\frac{c-k-\theta}{c-k+\theta}+\frac{-c-k-\theta}{-c-k+\theta}\right),\\
\tilde{B}(\theta|c) & =\frac{1}{2}\left(\frac{c-k-\theta}{c-k+\theta}-\frac{-c-k-\theta}{-c-k+\theta}\right),\\
k & =-i\frac{\pi}{2}\frac{n-4}{n-2}.
\end{align*}
For the classical limit we define a scaling variable $h$ for which
\begin{align*}
\theta & =\lambda/h, & c & =i/(ha).
\end{align*}
The classical limit is $h\to0$. In this limit the $R$-matrices are
proportional to the $\kappa$ matrices: 
\[
\lim_{h\to0}R_{i}(\lambda/h|i/(ha))\sim\kappa_{i}(\lambda|a).
\]

\subsubsection{Lagrangian and symmetries}

In the previous subsection we found reflection matrices parameterized
as \eqref{eq0} which leads to the following boundary condition: 
\begin{equation}
J_{1}^{R}=\frac{1}{2}\left[M,J_{0}^{R}\right].\label{bcPCMR}
\end{equation}
Using the left currents this condition takes the form: 
\begin{equation}
J_{1}^{L}=\frac{1}{2}\left[gMg^{-1},J_{0}^{L}\right].\label{bcPCML}
\end{equation}
One can obtain the same boundary condition in the Lagrangian description.
The Lagrangian density of the bulk theory is 
\[
\mathcal{L}_{PCM}=-\frac{1}{4}\Tr\left[\Jc^{L}\wedge*\Jc^{L}\right]=-\frac{1}{4}\Tr\left[\Jc^{R}\wedge*\Jc^{R}\right]
\]
Thus if we add a boundary Lagrangian function as 
\begin{equation}
L_{b}=\frac{1}{4}\Tr\left[MJ_{0}^{R}\right]\Big|_{x=0}\label{lagPCM}
\end{equation}
we get the boundary condition \eqref{bcPCMR}. This boundary condition
was already investigated in \citep{Moriconi:2001xz} and \citep{Aniceto:2017jor}.
It was shown that this is a conformal boundary condition for all $M\in\mathfrak{g}$.
Now we have just shown that it has a zero curvature representation
too for some special $M$s which satisfy the conditions \eqref{condMN}.

Now let us continue with the residual symmetries. The bulk Lagrangian
has $G_{L}\times G_{R}$ symmetries which are the left/right multiplications
with a constant group element: $g(x)\rightarrow g_{L}g(x)$ and $g(x)\rightarrow g(x)g_{R}$.
The transformations of the currents are the following: 
\begin{align*}
g_{L}: &  & \Jc^{L} & \rightarrow g_{L}\Jc^{L}g_{L}^{-1}, & \Jc^{R} & \rightarrow\Jc^{R},\\
g_{R}: &  & \Jc^{L} & \rightarrow\Jc^{L}, & \Jc^{R} & \rightarrow g_{R}^{-1}\Jc^{R}g_{R}.
\end{align*}
We can see that the boundary Lagrangian breaks the $G_{R}$ symmetry.
The remaining symmetry is $H_{R}<G_{R}$ where $H_{R}=\mathrm{exp}(\hn)$.
Since the current $\Jc^{R}$ is invariant under $G_{L}$, the $G_{L}$
symmetry is unbroken therefore the residual symmetry is $G_{L}\times H_{R}$.

One can derive the Noether charges by the variation of the action
but there is an easier way. We know that the $\Jc^{L}$ and $\Jc^{R}$
are the Noether currents of the bulk $G_{L}$ and $G_{R}$ symmetries.
Let us define the following charges: 
\begin{align*}
Q_{L} & =\int_{-\infty}^{0}J_{0}^{L}\di x,\\
Q_{R} & =\int_{-\infty}^{0}J_{0}^{R}\di x.
\end{align*}
By taking their time derivatives we obtain
\begin{align*}
\dot{Q}_{L} & =\int_{-\infty}^{0}\partial_{1}J_{1}^{L}\di x=J_{1}^{L}\Big|_{x=0}=\frac{1}{2}[gMg^{-1},J_{0}^{L}]\Big|_{x=0}=\frac{1}{2}\partial_{0}\left(gMg^{-1}\right)\Big|_{x=0}\\
\dot{Q}_{R} & =\int_{-\infty}^{0}\partial_{1}J_{1}^{R}\di x=J_{1}^{R}\Big|_{x=0}
\end{align*}
We can see that 
\begin{align}
\tilde{Q}_{L} & =Q_{L}-\frac{1}{2}\left(gMg^{-1}\right)\Big|_{x=0}\quad\text{and}\label{QL}\\
\tilde{Q}_{R} & =\Pi_{\hn}\left(Q_{R}\right)\label{QR}
\end{align}
are conserved charges.

Finally we note that we could have used the left current $\Jc^{L}$
with the $\kappa$-matrix 
\[
\kappa_{L}(\lambda)\sim1+\lambda M+\lambda^{2}N
\]
This implies that the right reflection matrix, the boundary condition
and the boundary Lagrangian are
\begin{align*}
\kappa_{R}(\lambda) & \sim1+\frac{1}{\lambda}g^{-1}Mg+\frac{1}{\lambda^{2}}g^{-1}Ng\\
J_{1}^{L} & =\frac{1}{2}[M,J_{0}^{L}]\\
L_{b} & =\frac{1}{4}\Tr[MJ_{0}^{L}]\Big|_{x=0}
\end{align*}
Therefore, in this case the residual symmetry is $H_{L}\times G_{R}$. 

\section{$\On{N}$ sigma model on the half line\label{sec:-sigma-model}}

The new reflection matrices of the PCM can be used to find new ones
for the $\On{N}$ sigma model. In particular, using the equivalence
between $\SU{2}$ PCM and the $\On{4}$ sigma model we have immediately
new reflection matrices for the $\On{N}$ sigma model when $N=4$.
This solution then can be generalized for even $N$.

\subsection{Lax formalism for the $\On{N}$ sigma model}

The field variables are $\n:\Sigma\rightarrow\mathbb{R}^{N}$ with
the $\n^{T}\n=1$ constrain. The bulk Lagrangian is 
\[
\mathcal{L}_{NL\sigma}=\frac{1}{2}\dd\n^{T}\wedge*\dd\n-\frac{1}{2}\sigma(\n^{T}\n-1).
\]
from which equation of motion follows: 
\[
\dd*\dd\n+(\dd\n^{T}\wedge\dd\n)\n=0.
\]
We can define an $\On{N}$ group element as: $h=1-2\n\n^{T}$ which
satisfies the following identities: $h^{T}h=1$ and $h=h^{T}$. Using
this, one can define a current: $\Ic=h\dd h=2\n\dd\n^{T}-2\dd\n\n^{T}$
which is the Noether current of the bulk global $\SO{N}$ symmetry.
The e.o.m with this current is $\dd*\Ic=0$ and the Lagrangian is
\[
\mathcal{L}_{NL\sigma}=-\frac{1}{16}\Tr\left[\Ic\wedge*\Ic\right].
\]
The Lax connection is very similar to the PCM but here the current
is constrained. 
\[
\hat{\Lc}(\lambda)=\frac{1}{1-\lambda^{2}}\Ic+\frac{\lambda}{1-\lambda^{2}}(*\Ic).
\]
The double row monodromy matrix can be defined similarly as it was
in PCMs. In the following we look for solutions of the boundary flatness
equation
\[
\kappa(\lambda)\hat{\mathcal{M}}(\lambda)-\hat{\mathcal{M}}(-\lambda)\kappa(\lambda)=\dot{\kappa}(\lambda),
\]

Let us start with the constant $\kappa$-matrices i.e. $\kappa(\lambda)=U$
where $U\in\mathrm{O}(N)$ therefore the boundary flatness equation
looks like
\[
U\left(\hat{J}_{0}-\lambda\hat{J}_{1}\right)-\left(\hat{J}_{0}+\lambda\hat{J}_{1}\right)U=0
\]
 which implies the following: 
\begin{align}
\lambda^{0} & : & \hat{J}_{0} & =U\hat{J}_{0}U^{-1},\label{eq:OkONk0}\\
\lambda^{1} & : & -\hat{J}_{1} & =U\hat{J}_{1}U^{-1}.\label{eq:OkONk1}
\end{align}

In this subsection, we assume that $U^{2}=\pm1$ but we do not derive
that. We will return to this at the next section. There are two kinds
of $U$s:
\begin{enumerate}
\item $U=\mathrm{diag}(1,\dots,1,-1,\dots,-1),$
\item $U=\left(\begin{array}{cc}
0_{n\times n} & -1_{n\times n}\\
1_{n\times n} & 0_{n\times n}
\end{array}\right)$ where $n=N/2$.
\end{enumerate}
Let us start with the first case. Let the number of $+1$s and $-1$s
be $N-k$ and $k$ respectively. Let us use the notation: $\n=\nt+\nh$,
with 
\[
\nt=(n_{1},\dots,n_{N-k},0,\dots,0)\quad,\qquad\nh=(0,\dots,0,n_{N-k+1},\dots,n_{N}).
\]
Using this, the equation \eqref{eq:OkONk0} is equivalent to
\[
\nt\dot{\nh}^{T}=\dot{\nt}\nh^{T}.
\]
Multiplying by $\nh$ from the right and $\nt^{T}$from the left,
we can obtain the following two equations
\begin{align}
\left(\nh^{T}\nh\right)\dot{\nt} & =\left(\nh^{T}\dot{\nh}\right)\nt,\label{eq:ontemp1}\\
\left(\nt^{T}\nt\right)\dot{\nh} & =\left(\nt^{T}\dot{\nt}\right)\nh.\label{eq:ontemp2}
\end{align}
Similarly, from \eqref{eq:OkONk1} we can get
\begin{align}
\left(\nh^{T}\nh\right)\nh' & =\left(\nh^{T}\nh'\right)\nh,\label{eq:ontemp1-1}\\
\left(\nt^{T}\nt\right)\nt' & =\left(\nt^{T}\nt'\right)\nt.\label{eq:ontemp2-1}
\end{align}

Let us assume that $\nh^{T}\nh=0$ which is equivalent to $\nt^{T}\nt=1$
and $\nh=0$. From this, the equations \eqref{eq:ontemp1} and \eqref{eq:ontemp1-1}
are satisfied trivially and the equations \eqref{eq:ontemp2} and
\eqref{eq:ontemp2-1} look like
\begin{align*}
\dot{\nh} & =0,\\
\nt' & =0,
\end{align*}
where we used that $0=\n^{T}\n=\nh^{T}\nh'+\nt^{T}\nt'=\nt^{T}\nt'$.
We can see that this is the restricted boundary condition to a sphere
$S^{k}$ with maximal radius. Analogously, if we assume that $\nt^{T}\nt=0$
then
\begin{align*}
\dot{\nt} & =0,\\
\nh' & =0,
\end{align*}
which is the restricted bc to $S^{N-k}$ with maximal radius.

What happens when $\nh^{T}\nh\neq0$ and $\nt^{T}\nt\neq0$. Let us
multiply \eqref{eq:ontemp1} with $\nt^{T}$ form the left:
\[
\left(\nh^{T}\nh\right)\left(\nt^{T}\dot{\nt}\right)=\left(\nh^{T}\dot{\nh}\right)\left(\nt^{T}\nt\right),
\]
Using that $\left(\nh^{T}\dot{\nh}\right)+\left(\nt^{T}\dot{\nt}\right)=0$
\[
0=\left(\nh^{T}\nh+\nt^{T}\nt\right)\left(\nt^{T}\dot{\nt}\right)=\left(\nt^{T}\dot{\nt}\right)
\]
therefore $\nt^{T}\dot{\nt}=\nh^{T}\dot{\nh}=0$ which implies
\begin{align*}
\dot{\nt} & =0,\\
\dot{\nh} & =0.
\end{align*}
From this and equations \eqref{eq:ontemp1-1}, \eqref{eq:ontemp2-1},
we can see that there are too many boundary conditions therefore the
$\hat{J}_{0}=U\hat{J}_{0}U^{-1}$ and $\hat{J}_{1}=-U\hat{J}_{1}U^{-1}$
are consistent boundary conditions if and only if $\nh=0$ or $\nt=0$.
In Subsection \eqref{subsec:Poisson-bracket-in} we will see that
$\kappa=\mathrm{diag}(1,\dots,1,-1,\dots,-1)$ satisfies the classical
boundary Yang-Baxter equation if and only if $\nh=0$ or $\nt=0$.

Let us continue with the second case i.e. $U^{\text{T}}=-U$. Let
us start with equation \eqref{eq:ontemp2}:
\[
\n\n'^{T}-\n'\n^{T}=U\n\n'^{T}U-U\n'\n^{T}U
\]
Let us multiply this with $\n$ from the right:
\[
\n'=U\n\left(\n^{T}U\n'\right)
\]
From this we can obtain the following two equations
\begin{align*}
\n'\n^{T} & =+U\n\n^{T}\left(\n^{T}U\n'\right)\\
\n\n'^{T} & =-\n\n^{T}U\left(\n^{T}U\n'\right)
\end{align*}
therefore 
\[
J_{1}=-2\left(\n\n^{T}U+U\n\n^{T}\right)\left(\n^{T}U\n'\right)
\]
Let us multiply this with $U$ from the left and $U^{T}$ from the
right.

\[
UJ_{1}U^{T}=-2\left(U\n\n^{T}+\n\n^{T}U\right)\left(\n^{T}U\n'\right)=J_{1}
\]
Using this and the original equation \eqref{eq:ontemp2} we can obtain
that $J_{1}=0$ which is equivalent to $\n'=0$. But we also have
equation \eqref{eq:ontemp1} therefore we have too many boundary condition
which means that $\hat{J}_{0}=U\hat{J}_{0}U^{-1}$ and $\hat{J}_{1}=-U\hat{J}_{1}U^{-1}$
are not consistent boundary conditions at the second case. We will
also see at Subsection \eqref{subsec:Poisson-bracket-in} that the
$\kappa$-matrix of the second case do not satisfy the classical boundary
Yang-Baxter equation.

\subsection{Spectral parameter dependent solution for $N=4$}

In the last section, we found a new spectral parameter dependent reflection
matrix for the $\mathrm{SU}(2)$ PCM. Since this model is equivalent
to the $\On{4}$ sigma model we can obtain a new non-constant $\kappa$-matrix
for the $\On{4}$ sigma model by changing the notation to the $\On{4}$
sigma model language. We will see that this is a \textit{spectral
parameter and field (!) dependent} reflection matrix. 

Thus we need to develop a dictionary between the $\mathrm{SU}(2)$
PCM and the $\On{4}$ sigma model. Let us introduce the following
tensor: 
\begin{align*}
\sigma_{\alpha\dal}^{1} & =\begin{pmatrix}0 & i\\
i & 0
\end{pmatrix}, & \sigma_{\alpha\dal}^{2} & =\begin{pmatrix}0 & 1\\
-1 & 0
\end{pmatrix}, & \sigma_{\alpha\dal}^{3} & =\begin{pmatrix}i & 0\\
0 & -i
\end{pmatrix}, & \sigma_{\alpha\dal}^{4} & =\begin{pmatrix}1 & 0\\
0 & 1
\end{pmatrix}.
\end{align*}
which satisfies the following relations: 
\begin{align*}
\sigma_{\alpha\dal}^{i}\bsig_{i}^{\beta\dbe} & =2\delta_{\alpha}^{\beta}\delta_{\dal}^{\dbe},\\
\sigma_{\alpha\dal}^{i}\bsig_{j}^{\alpha\dal} & =2\delta_{i}^{j},
\end{align*}
where $\bsig_{i}^{\alpha\dal}$ is the complex conjugate of $\sigma_{\alpha\dal}^{i}$.
Using this we can change the basis in which the group element $g_{4}=\SO{4}$
is factorized. 
\begin{align*}
\frac{1}{2}\sigma_{\alpha\dal}^{i}(g_{4})_{i}^{j}\bsig_{j}^{\beta\dbe} & =(g_{L})_{\alpha}^{\beta}(g_{R})_{\dal}^{\dbe}, & (g_{4})_{i}^{j} & =\frac{1}{2}\bsig_{i}^{\alpha\dal}(g_{L})_{\alpha}^{\beta}(g_{R})_{\dal}^{\dbe}\sigma_{\beta\dbe}^{j}.
\end{align*}
In this basis: 
\begin{equation}
\n=g_{4}\n_{0}\quad\rightarrow\quad\n=g_{L}g_{R}^{T}=\begin{pmatrix}n_{4}+in_{3} & in_{1}+n_{2}\\
in_{1}-n_{2} & n_{4}-in_{3}
\end{pmatrix}=g\in\SU{2},\label{gdef}
\end{equation}
if $\n_{0}=(0,0,0,1)$. 

We can also find the relation between the variables of the $\On{4}$
model ($h,\Ic$) and the $\SU{2}$ PCM ($g,\Jc^{L/R}$). Using $\n=g_{4}\n_{0}$
and $h=1-2\n\n^{T}$ we obtain that $h=g_{4}jg_{4}^{T}$ where $j=1-2\n_{0}\n_{0}^{T}=\mathrm{diag}(1,1,1,-1)\in\On{4}$.
Since $\mathrm{det}(j)=-1$, $j$ is not factorized in the new basis:
\[
j\rightarrow(\sigma_{2}\otimes\sigma_{2}^{\dagger})P,
\]
where $P$ is the permutation operator.

The group element $h$ in the new basis takes the form: 
\[
h=(g_{L}\otimes g_{R})(\sigma_{2}\otimes\sigma_{2}^{\dagger})P(g_{L}^{\dagger}\otimes g_{R}^{\dagger})=((g\sigma_{2})\otimes(g\sigma_{2})^{\dagger})P=P((g\sigma_{2})^{\dagger}\otimes(g\sigma_{2})).
\]
($g$ was defined in \eqref{gdef}) In the last line we used the following
property: $\sigma_{2}g\sigma_{2}^{\dagger}=\bar{g}$ and $\bar{g}$
denotes the complex conjugate of $g$. We can see that $h$ is not
factorized. This is because $h$ is not an element of $\SO{4}$. It
is convenient to introduce a new notation: 
\[
h_{2}=g\sigma_{2},\quad\rightarrow\quad h=h_{2}\otimes h_{2}^{\dagger}P.
\]
Let us calculate $\Ic$ in the new basis. 
\begin{align}
\Ic=h\dd h=\Jc^{L}\otimes1+1\otimes\bar{\Jc}^{R},\label{eqIc}
\end{align}
where $\bar{\Jc}^{R}$ denotes the complex conjugate of $\Jc^{R}$.
The Lax connection in the new basis is: 
\[
\hat{\Lc}(\lambda)=\left(\frac{1}{1-\lambda^{2}}\Jc^{L}+\frac{\lambda}{1-\lambda^{2}}*\Jc^{L}\right)\otimes1+1\otimes\left(\frac{1}{1-\lambda^{2}}\bar{\Jc}^{R}+\frac{\lambda}{1-\lambda^{2}}*\bar{\Jc}^{R}\right)=\Lc^{L}(\lambda)\otimes1+1\otimes\bar{\Lc}^{R}(\lambda).
\]
Therefore the monodromy matrix of the $\On{4}$ sigma model factorized
in the following way: 
\[
\hat{T}(\lambda)=T_{L}(\lambda)\otimes\bar{T}_{R}(\lambda).
\]
The double row monodromy matrix in the new basis reads: 
\[
\hat{\Omega}(\lambda)=(T_{L}(-\lambda)^{-1}\otimes\bar{T}_{R}(-\lambda)^{-1})\kappa_{4}(\lambda)(T_{L}(\lambda)\otimes\bar{T}_{R}(\lambda)).
\]

Before we calculate the new $\kappa$-matrix let us apply the formula
above to the known constant reflection matrices. The simplest known
$\kappa_{4}$ is the identity matrix. This is factorized in the spinor
basis: $\kappa^{L}=\kappa^{R}=1$. Another known reflection matrix
is $\kappa=\mathrm{diag}(-1,-1,1,1)$ in the vector basis. If we change
the basis we get: 
\[
\kappa=\begin{pmatrix}1 & 0 & 0 & 0\\
0 & -1 & 0 & 0\\
0 & 0 & -1 & 0\\
0 & 0 & 0 & 1
\end{pmatrix}=\begin{pmatrix}1 & 0\\
0 & -1
\end{pmatrix}\otimes\begin{pmatrix}1 & 0\\
0 & -1
\end{pmatrix}
\]
thus $\kappa^{R}=\kappa^{L}=\mathrm{diag}(1,-1)$. These two reflection
factors are consistent if they satisfy the inversion property \eqref{eq:invKappa}
i.e.
\begin{align*}
\kappa_{L}(\lambda) & =g(0)\kappa_{R}(1/\lambda)g^{\dagger}(0)
\end{align*}
which means that $g$ has to commute with them therefore $g$ is restricted
to $H=\mathrm{U}(1)$ at the boundary.

There is another known reflection matrix: $\kappa=\mathrm{diag}(1,1,1,-1)$
in the vector basis. If we change the basis we get: 
\[
\kappa=\begin{pmatrix}0 & 0 & 0 & -1\\
0 & 1 & 0 & 0\\
0 & 0 & 1 & 0\\
-1 & 0 & 0 & 0
\end{pmatrix}=(\sigma_{2}\otimes\sigma_{2}^{\dagger})P.
\]
We can see this matrix is not factorized. Using this formula for the
monodromy matrix, we obtain that 
\begin{multline*}
\hat{\Omega}(\lambda)=P(\bar{T}_{R}^{-1}(-\lambda)\otimes T_{L}^{-1}(-\lambda))(\sigma_{2}^{\dagger}\otimes\sigma_{2})(T_{L}(\lambda)\otimes\bar{T}_{R}(\lambda))=\\
=P(\sigma_{2}^{\dagger}T_{R}^{-1}(-\lambda)T_{L}(\lambda))\otimes(T_{L}^{-1}(-\lambda)T_{R}(\lambda)\sigma_{2}).
\end{multline*}
This theory is consistent in the principal model language if $g=g^{\dagger}$
at the boundary which is the boundary conditions \eqref{eq:adj}.

These were the relations of the well known reflection matrices of
the $\mathrm{SU}(2)$ PCM and the $\On{4}$ sigma model. Let us continue
with the new one. In the last section we found new reflection matrices
for the PCM model which for $\mathfrak{g}=\mathfrak{su}(2)$ simplifies
to
\[
\kappa^{R}(\lambda)\sim\left(1+\lambda M_{R}\right),
\]
where $M_{R}$ is an arbitrary element of $\su{2}$. Without loss
of generality one can choose $M_{R}=a\sigma_{2}$. We have seen that
$\kappa^{L}(\lambda)=g\kappa^{R}(1/\lambda)g^{\dagger}$ so we have
\begin{equation}
\kappa(\lambda)\sim\left(1+\frac{1}{\lambda}gM_{R}g^{\dagger}\right)\otimes\left(1+\lambda\bar{M}_{R}\right)=1\otimes1+\lambda1\otimes\bar{M_{R}}+\frac{1}{\lambda}(gM_{R}g^{\dagger})\otimes1+(gM_{R}g^{\dagger})\otimes\bar{M}_{R},\label{kappaO4}
\end{equation}
Let us denote $1\otimes\bar{M_{R}}$ in the vector representation
by $M$. In the spinor basis $hMh$ looks like
\begin{align}
hMh\rightarrow((g\sigma_{2})\otimes(g\sigma_{2})^{\dagger})P(1\otimes\bar{M}_{R})P((g\sigma_{2})^{\dagger}\otimes(g\sigma_{2}))=(gM_{R}g^{\dagger})\otimes1,\label{hMh}
\end{align}
therefore 
\[
MhMh=hMhM=\frac{1}{2}\left[M,hMh\right]_{+}\rightarrow(gM_{R}g^{\dagger})\otimes\bar{M}_{R}
\]
Based on the above formulas, the new $\kappa$-matrix for $\On{4}$
takes the following form: 
\begin{equation}
\kappa(\lambda)\sim1+\lambda M+\frac{1}{\lambda}hMh+\frac{1}{2}\left[M,hMh\right]_{+},\label{kappaOn}
\end{equation}
where the matrix $M$ looks like 
\[
M=a\begin{pmatrix}0 & 0 & 1 & 0\\
0 & 0 & 0 & 1\\
-1 & 0 & 0 & 0\\
0 & -1 & 0 & 0
\end{pmatrix}.
\]
We can see that this $\kappa$ is spectral parameter and field dependent
too. We can give the boundary condition which correspond to this $\kappa$
from the boundary conditions of $\SU{2}$ PCM \eqref{bcPCMR},\eqref{bcPCML}
and \eqref{eqIc}. 
\[
\hat{J}_{1}=J_{1}^{L}\otimes1+1\otimes\bar{J}_{1}^{R}=\frac{1}{2}[gM_{R}g^{\dagger},J_{0}^{L}]\otimes1+\frac{1}{2}1\otimes[\bar{M}_{R},\bar{J}_{0}^{R}]
\]
Using the definition of $M$ 
\[
[M,\hat{J}_{0}]=[1\otimes\bar{M}_{R},J_{0}^{L}\otimes1+1\otimes\bar{J}_{0}^{R}]=1\otimes[\bar{M}_{R},\bar{J}_{0}^{R}]
\]
and using \eqref{hMh} 
\[
[hMh,\hat{J}_{0}]=[(gM_{R}g^{\dagger})\otimes1,J_{0}^{L}\otimes1+1\otimes\bar{J}_{0}^{R}]=[gM_{R}g^{\dagger},J_{0}^{L}]\otimes1
\]
Therefore the boundary condition in language of the $\On{4}$ model
is: 
\begin{equation}
\hat{J}_{1}=\frac{1}{2}[M+hMh,\hat{J}_{0}].\label{bcOnJ}
\end{equation}
This boundary condition was investigated in \citep{Aniceto:2017jor}.
Using the definition $\Ic=h\dd h=2\n\dd\n^{T}-2\dd\n\n^{T}$, we can
get an equivalent form : 
\begin{equation}
\n'=M\dot{\n}-(\n^{T}M\dot{\n})\n.\label{bcOnn}
\end{equation}
From the boundary Lagrangian of the $\SU{2}$ PCM we get 
\[
L_{b}=\frac{1}{4}\Tr[M_{R}J_{0}^{R}]=\frac{1}{8}\Tr[(1\otimes\bar{M}_{R})(J_{0}^{L}\otimes1+1\otimes\bar{J}_{0}^{R})],
\]
therefore 
\begin{equation}
L_{b}=\frac{1}{8}\Tr[M\hat{J}_{0}]\label{lagrangeOnJ}
\end{equation}
which agrees with \citep{Aniceto:2017jor}. Using the variables $\n$:
\begin{equation}
L_{b}=-\frac{1}{2}\n^{T}M\dot{\n}.\label{lagrangeOnn}
\end{equation}
Finally, we can see that the residual symmetry is $\mathrm{U}(2)\cong\SU{2}_{L}\times\mathrm{U}(1)_{R}$
which is a subgroup of $\SU{2}_{L}\times\SU{2}_{R}\cong\SO{4}$. We
saw in the PCMs that we have conserved charges $\tilde{Q}_{L}$ and
$\tilde{Q}_{R}$. The conserved charge in the $\SO{4}$ language are:
\[
\tilde{Q}=\tilde{Q}_{L}\otimes1+1\otimes\bar{\tilde{Q}}_{R}=Q_{L}\otimes1+1\otimes\overline{\Pi_{\hn_{R}}\left(Q_{R}\right)}-\frac{1}{2}(gM_{R}g^{\dagger})\Big|_{x=0}\otimes1.
\]
which is equivalent to 
\begin{equation}
\tilde{Q}=\Pi_{\hn}\left(Q\right)-\frac{1}{2}hMh\Big|_{x=0}=\Pi_{\hn}\left(Q-\frac{1}{2}hMh\Big|_{x=0}\right),\label{QO4}
\end{equation}
where $\hn=\su{2}_{L}\oplus\mathfrak{u}(1)_{R}$, and $Q$ is the
bulk part of the charge: 
\[
Q=\int_{-\infty}^{0}\hat{J}_{0}\dd x.
\]

\subsection{Generalization for $N=2n$}

The result for $N=4$ can be generalized for any even $N$. We assume
that equation \eqref{kappaOn} can be used as $\kappa$ matrix for
$N=2n$ i.e.
\begin{equation}
\kappa(\lambda)\sim1+\lambda M+\frac{1}{\lambda}hMh+\frac{1}{2}\left[M,hMh\right]_{+},\label{kappaOn-1}
\end{equation}
where 
\[
M=a\begin{pmatrix}0_{n\times n} & 1_{n\times n}\\
-1_{n\times n} & 0_{n\times n}
\end{pmatrix}.
\]
We have to prove that the time derivative of the double row monodromy
matrix is zero when the boundary condition is satisfied. The quantity
$\partial_{0}\hat{\Omega}$ is zero when the boundary flatness condition
is satisfied
\begin{equation}
\kappa(\lambda)\hat{\mathcal{M}}(\lambda)\Big|_{x=0}-\hat{\mathcal{M}}(-\lambda)\Big|_{x=0}\kappa(\lambda)=\dot{\kappa}(\lambda),\label{reflexEqOn}
\end{equation}
Now the RHS is not zero since the $\kappa$ has field dependence.
\[
\dot{\kappa}(\lambda)\sim\partial_{0}\left(1+\lambda M+\frac{1}{\lambda}hMh+\frac{1}{2}\left[M,hMh\right]_{+}\right)=\frac{1}{\lambda}\left[hMh,\hat{J}_{0}\right]+\frac{1}{2}\left[M,\left[hMh,\hat{J}_{0}\right]\right]_{+}.
\]
Using this, equation \eqref{reflexEqOn} leads to the following three
equations: 
\begin{align*}
\lambda^{0}: &  & \frac{1}{2}\left[\left[M,hMh\right]_{+},\hat{J}_{0}\right]-\left[hMh,\hat{J}_{1}\right]_{+} & =\frac{1}{2}\left[M,\left[hMh,\hat{J}_{0}\right]\right]_{+}\\
\lambda^{1}: &  & \left[M,\hat{J}_{0}\right]-2\hat{J}_{1}-\frac{1}{2}\left[\left[M,hMh\right]_{+},\hat{J}_{1}\right]_{+} & =-\left[hMh,\hat{J}_{0}\right]\\
\lambda^{2}: &  & -\left[M,\hat{J}_{1}\right]_{+} & =-\frac{1}{2}\left[M,\left[hMh,\hat{J}_{0}\right]\right]_{+}
\end{align*}
If we take the anti-commutator of the boundary condition \eqref{bcOnJ}
with $M$ then we will see that the third equation is satisfied. If
we use the following identity 
\[
\left[\left[M,hMh\right]_{+},\hat{J}_{0}\right]+\left[\left[\hat{J}_{0},M\right],hMh\right]_{+}-\left[\left[hMh,\hat{J}_{0}\right],M\right]_{+}=0
\]
then the first equation can be written as 
\[
\left[hMh,\hat{J}_{1}\right]_{+}=\frac{1}{2}\left[hMh,\left[M,\hat{J}_{0}\right]\right]_{+}.
\]
This is also follows from the boundary condition.

Only the second equation remained. We have to prove that the following
term vanish: 
\begin{equation}
\frac{1}{2}\left[\left[M,hMh\right]_{+},\hat{J}_{1}\right]_{+}\label{anticom1}
\end{equation}
Using the definition of $h$, we obtain that 
\[
MhMh=M(M-2\n\n^{T}M-2M\n\n^{T})=-a^{2}h-2M\n\n^{T}M=hMhM.
\]
Therefore 
\[
\frac{1}{2}\left[M,hMh\right]_{+}=-a^{2}h-2M\n\n^{T}M.
\]
Since $\hat{J}_{1}$ is anti-commuting with $h$ by definition, we
only have to prove only that $M\n\n^{T}M$ is anti-commuting with
$\hat{J}_{1}$ too. For this, we have to use the boundary condition
\eqref{bcOnJ} which can be written as 
\[
\hat{J}_{1}=-2M\dot{\n}\n^{T}-2\n\dot{\n}^{T}M
\]
Using this, we obtain that 
\[
\left[M\n\n^{T}M,\hat{J}_{1}\right]_{+}=\left[M\n\n^{T}M,-2M\dot{\n}\n^{T}-2\n\dot{\n}^{T}M\right]_{+}=0.
\]
Therefore the expression \eqref{anticom1} is vanishing so the second
equation is satisfied too which implies that the double row monodromy
matrix is conserved if the boundary condition \eqref{bcOnJ} is satisfied.

After this derivation, let us continue with the symmetries. Now the
residual symmetry is $\mathrm{U}(n)<\SO{2n}$ where $H=\mathrm{U}(n)$
is the subgroup which commutes with $M$. Since $\SO{2n}/\mathrm{U}(n)$
is a symmetric space we have a $\mathbb{Z}_{2}$ graded decomposition
$\so{2n}=\hn\oplus\fn$ where $\hn=$ is the Lie-algebra of $\mathrm{U}(n)$
so $\hn=\su{n}\oplus\ue$. The $\ue$ is generated by $M$ so $[M,\hn]=0$
and $[M,\fn]\subset\hn$ therefore 
\begin{equation}
[M,X]\in\fn,\label{comM}
\end{equation}
for any $X\in\so{2n}$.

For conserved charges, we can generalize the formula \eqref{QO4}.
\[
\tilde{Q}=\Pi_{\hn}\left(Q-\frac{1}{2}hMh\Big|_{x=0}\right)=\Pi_{\hn}\left(\int_{-\infty}^{0}\hat{J}_{0}\dd x-\frac{1}{2}hMh\Big|_{x=0}\right).
\]
We can check the conservation of these charges. 
\[
\dot{\tilde{Q}}=\Pi_{\hn}\left(\dot{Q}-\frac{1}{2}(\dot{h}Mh+hM\dot{h})\Big|_{x=0}\right)=\Pi_{\hn}\left(\hat{J}_{1}-\frac{1}{2}\left[hMh,\hat{J}_{0}\right]\right)\Big|_{x=0}=\frac{1}{2}\Pi_{\hn}\left[M,\hat{J}_{0}\right]\Big|_{x=0}=0,
\]
where we used \eqref{comM}.

The boundary Lagrangian can be written in the same form as we had
for the case $N=4$ \eqref{lagrangeOnJ} or \eqref{lagrangeOnn}:
\[
L_{b}=\frac{1}{8}\Tr[M\hat{J}_{0}]=-\frac{1}{2}\n^{T}M\dot{\n}.
\]
 These have been studied earlier in \citep{Aniceto:2017jor} where
it was showed that this is a conform boundary condition for any $M\in\mathfrak{so}(2n)$
but in this paper we showed more, namely that it has a zero curvature
representation only when $M^{2}\sim$1.

\section{Poisson algebra of double row monodromy matrices\label{sec:Poisson-algebra-of}}

In the previous sections we found new zero curvature representation
of PCMs and $\On{N}$ sigma models on a half line. This implies the
existence of infinitely many conserved charges. In this section we
want to prove that these conserved charges are in involution. For
this we determine the Poisson algebra of the double row monodromy
matrices (whose trace is the generating function of these charges).
In the first subsection we summarize the formulas of general ``bulk''
non-ultralocal theories based on \citep{Maillet:1985ek}. After that
we derive the Poisson-algebra of the double row monodromy matrices
and their consistency condition (which is the classical boundary Yang-Baxter
equation) when the Poisson-bracket of the reflection matrix and the
Lax-connection is not zero. This is a new result because, so far Poisson-algebras
of non-ultralocal theories with boundaries were investigated only
when the $\kappa$-matrix was field independent \citep{Corrigan:1996nt,Mann:2006rh}. 

In the second and the third subsection we apply these general formulas
for PCMs and non linear sigma models. We will use the following notations:
\begin{align*}
X_{1} & =X\otimes1 & X_{2} & =1\otimes X\\
Y_{12} & =Y\otimes1 & Y_{23} & =1\otimes Y
\end{align*}
where $X\in\mathrm{End}(V)$ and $Y\in\mathrm{End}(V)\otimes\mathrm{End}(V)$
for a vector space $V$.

\subsection{The double-row monodromy matrices of non-ultralocal theories}

The general Poisson-brackets of the space-like components of the Lax-connection
for non-ultralocal theories are the following \citep{Maillet:1985ek}:
\begin{align}
\{\mathcal{L}_{1}(x|\lambda_{1}),\mathcal{L}_{2}(y|\lambda_{2})\}=- & \bigl[r_{12}(x|\lambda_{1},\lambda_{2}),\mathcal{L}_{1}(x|\lambda_{1})+\mathcal{L}_{2}(x|\lambda_{2})\bigr]\delta(x-y)+\nonumber \\
+ & \bigl[s_{12}(x|\lambda_{1},\lambda_{2}),\mathcal{L}_{1}(x|\lambda_{1})-\mathcal{L}_{2}(x|\lambda_{2})\bigr]\delta(x-y)-\nonumber \\
- & \left(r_{12}(x|\lambda_{1},\lambda_{2})+s_{12}(x|\lambda_{1},\lambda_{2})-r_{12}(y|\lambda_{1},\lambda_{2})+s_{12}(y|\lambda_{1},\lambda_{2})\right)\delta'(x-y),\label{eq:LPalg}
\end{align}
From the anti-symmetry of the Poisson bracket \eqref{eq:poissonT}
we obtain the following constraints on $r$- and $s$-matrices:
\begin{align*}
r_{12}(\lambda_{1},\lambda_{2}) & =-r_{21}(\lambda_{2},\lambda_{1}),\\
s_{12}(\lambda_{1},\lambda_{2}) & =+s_{21}(\lambda_{2},\lambda_{1}).
\end{align*}
We can generalize the one row monodromy matrix for general paths from
$y$ to $x$:
\[
T(x,y|\lambda)=\Pexp\left(-\int_{y}^{x}\mathcal{L}(z|\lambda)\dd z\right).
\]
Let $x_{1},x_{2},y_{1},y_{2}$ be different positions and $x_{1,2}>y_{1,2}$
then the general non-ultralocal Poisson-brackets of the monodromy
matrices are the following \citep{Maillet:1985ek}:
\begin{align}
\{T_{1}(x_{1},y_{1}|\lambda_{1}),T_{2}(x_{2},y_{2}|\lambda_{2})\} & =t_{12}^{-}\left(R_{12}^{-}t_{12}-t_{12}R_{12}^{+}\right)t_{12}^{+}.\label{eq:poissonT}
\end{align}
where $x_{0}=min(x_{1},x_{2})$, $y_{0}=max(y_{1},y_{2})$ and
\begin{align*}
t_{12}^{-} & =T_{1}(x_{1},x_{0}|\lambda_{1})T_{2}(x_{2},x_{0}|\lambda_{2})\\
t_{12} & =T_{1}(x_{0},y_{0}|\lambda_{1})T_{2}(x_{0},y_{0}|\lambda_{2})\\
t_{12}^{+} & =T_{1}(y_{0},y_{1}|\lambda_{1})T_{2}(y_{0},y_{2}|\lambda_{2})\\
R_{12}^{-} & =r_{12}(x_{0}|\lambda_{1},\lambda_{2})+\mathrm{sgn}(x_{1}-x_{2})s_{12}(x_{0}|\lambda_{1},\lambda_{2})\\
R_{12}^{+} & =r_{12}(y_{0}|\lambda_{1},\lambda_{2})+\mathrm{sgn}(y_{2}-y_{1})s_{12}(y_{0}|\lambda_{1},\lambda_{2})
\end{align*}
This Poisson-bracket satisfies the Jacobi identity (for not coinciding
points) if the generalized classical Yang-Baxter equation is satisfied:
\begin{multline*}
\left[r_{23}(\lambda_{2},\lambda_{3})+s_{23}(\lambda_{2},\lambda_{3}),r_{13}(\lambda_{1},\lambda_{3})+s_{13}(\lambda_{1},\lambda_{3})\right]+\\
+\left[r_{23}(\lambda_{2},\lambda_{3})+s_{23}(\lambda_{2},\lambda_{3}),r_{12}(\lambda_{1},\lambda_{2})+s_{12}(\lambda_{1},\lambda_{2})\right]+\\
+\left[r_{13}(\lambda_{1},\lambda_{3})+s_{13}(\lambda_{1},\lambda_{3}),r_{12}(\lambda_{1},\lambda_{2})-s_{12}(\lambda_{1},\lambda_{2})\right]+\\
+H_{123}^{(r+s)}(\lambda_{1},\lambda_{2},\lambda_{3})-H_{213}^{(r+s)}(\lambda_{2},\lambda_{1},\lambda_{3})=0
\end{multline*}
where
\[
\left\{ \mathcal{L}_{1}(x|\lambda_{1}),\left(r_{23}(y|\lambda_{2},\lambda_{3})+s_{23}(y|\lambda_{2},\lambda_{3})\right)\right\} =-H_{123}^{(r+s)}(\lambda_{1},\lambda_{2},\lambda_{3})\delta(x-y).
\]

For the calculation of the Poisson bracket of the global monodromy
matrices \eqref{eq:globT} we have to take the limits $x_{1}\to x_{2}$
and $y_{1}\to y_{2}$. However, the Poisson bracket \eqref{eq:poissonT}
is not continuous due to the non ultra-locality. It is obvious that
the equal intervals limit of the canonical brackets does not exist
in a strong sense. More precisely, any strong definition implies the
breakdown of the Jacobi identity for the canonical brackets of the
global monodromy matrices \eqref{eq:globT}.

However, it is possible to define this limit in a weak sense with
respect to the canonical brackets based on a split-point procedure
and a generalized symmetric limit. We consider canonical brackets
of several monodromy matrices defined on intervals having coinciding
end points. In order to compute them, let us first split the coinciding
points and use \eqref{eq:poissonT} which then gives a completely
consistent expression. Then if we symmetrize on all the possible splittings
and go to the limit of equal points we get the ``weak'' algebras
e.g. the weak algebra of the global monodromy matrices:
\[
\left\{ T_{1}(\lambda_{1}),T_{2}(\lambda_{2})\right\} =r_{12}(0|\lambda_{1},\lambda_{2})T_{1}(\lambda_{1})T_{2}(\lambda_{2})-T_{1}(\lambda_{1})T_{2}(\lambda_{2})r_{12}(-\infty|\lambda_{1},\lambda_{2}).
\]
The formulas above can be found in \citep{Maillet:1985ek} but in
this paper we use a different conventions for the Lax-pair i.e. we
have to change $\mathcal{L}\to-\mathcal{L}$ to get the formulas in
\citep{Maillet:1985ek}. In the following we derive the Poisson-algebra.
For this we need the $\kappa$-matrices which were derived in the
previous sections. We saw that these matrices can depend on the fields
but do not on the derivative of the fields therefore we assume that
\[
\left\{ \mathcal{L}_{1}(x|\lambda_{1}),\kappa_{2}(\lambda_{2})\right\} =-G_{12}(\lambda_{1},\lambda_{2})\delta(x).
\]
Let us continue with the generalized double row monodromy matrix:
\[
\Omega(x|\lambda)\coloneqq T^{-1}(0,x|-\lambda)\kappa(\lambda)T(0,x|\lambda)=T(x,0|-\lambda)\kappa(\lambda)T(0,x|\lambda).
\]
The Poisson bracket of $\Omega(x|\lambda)$ and $\Omega(y|\mu)$ are
not well defined even when $x\neq y$ therefore we have to use the
split-point procedure. For this, we can define a shifted double row
monodromy matrix:
\[
\Omega^{\Delta}(x|\lambda)=T(x,\Delta|-\lambda)\kappa(\Delta|\lambda)T(\Delta,x|\lambda)
\]
where $\Delta<0$. A general $\kappa$-matrix depends on the boundary
value of the fields $\phi_{a}(0)$ (i.e. $\kappa(\lambda)=\kappa(\phi_{a}(0)|\lambda)$)
but we can extend this to arbitrary space coordinate:
\[
\kappa(\Delta|\lambda)=\kappa(\phi_{a}(\Delta)|\lambda).
\]
Using these the Poisson bracket of monodromy matrices are
\[
\{\Omega_{1}(x_{1}|\lambda_{1}),\Omega_{2}(x_{2}|\lambda_{2})\}\coloneqq\frac{1}{2}\lim_{\Delta\to0}\left[\{\Omega_{1}(x_{1}|\lambda_{1}),\Omega_{2}^{\Delta}(x_{2}|\lambda_{2})\}+\{\Omega_{1}^{\Delta}(x_{1}|\lambda_{1}),\Omega_{2}(x_{2}|\lambda_{2})\}\right].
\]
In the following we assume that 
\begin{align*}
r(-\lambda_{1},-\lambda_{2}) & =-r(\lambda_{1},\lambda_{2}),\\
s(-\lambda_{1},-\lambda_{2}) & =-s(\lambda_{1},\lambda_{2}).
\end{align*}
Now we can calculate the symmetric limit: 
\begin{align}
\{\Omega_{1}(x_{1}|\lambda_{1}),\Omega_{2}(x_{2}|\lambda_{2})\}= & t_{12}^{-}\left(\left[R_{12},\omega_{12}\right]+\omega_{1}^{(1)}\tilde{R}_{12}\omega_{2}^{(2)}-\omega_{2}^{(2)}\tilde{R}_{12}\omega_{1}^{(2)}\right)t_{12}^{+}-\nonumber \\
- & T_{12}^{-}\biggl(\bigl[r_{12}(0|\lambda_{1},\lambda_{2}),\kappa_{1}(\lambda_{1})\kappa_{2}(\lambda_{2})\bigr]+\nonumber \\
 & \qquad+\kappa_{1}(\lambda_{1})r_{12}(0|\lambda_{1},-\lambda_{2})\kappa_{2}(\lambda_{2})-\kappa_{2}(\lambda_{2})r_{12}(0|\lambda_{1},-\lambda_{2})\kappa_{1}(\lambda_{1})+\nonumber \\
 & \qquad+\frac{1}{2}\Bigl(G_{12}(-\lambda_{1},\lambda_{2})\kappa_{1}(\lambda_{1})-\kappa_{1}(\lambda_{1})G_{12}(\lambda_{1},\lambda_{2})-\nonumber \\
 & \qquad-G_{21}(-\lambda_{2},\lambda_{1})\kappa_{2}(\lambda_{2})+\kappa_{2}(\lambda_{2})G_{21}(\lambda_{2},\lambda_{1})\Bigr)\biggr)T_{12}^{+}.
\end{align}
where $x_{0}=max(x_{1},x_{2})$ and
\begin{align*}
t_{12}^{-} & =T_{1}(x_{1},x_{0}|-\lambda_{1})T_{2}(x_{2},x_{0}|-\lambda_{2})\\
\omega_{12} & =\Omega_{1}(x_{0}|\lambda_{1})\Omega_{2}(x_{0}|\lambda_{2})\\
\omega^{(1)} & =\Omega(x_{0}|\lambda_{1})\\
\omega^{(2)} & =\Omega(x_{0}|\lambda_{2})\\
t_{12}^{+} & =T_{1}(x_{0},x_{1}|\lambda_{1})T_{2}(x_{0},x_{2}|\lambda_{2})\\
R_{12} & =r_{12}(x_{0}|\lambda_{1},\lambda_{2})+\mathrm{sgn}(x_{2}-x_{1})s_{12}(x_{0}|\lambda_{1},\lambda_{2})\\
\tilde{R}_{12} & =r_{12}(x_{0}|\lambda_{1},-\lambda_{2})+\mathrm{sgn}(x_{2}-x_{1})s_{12}(x_{0}|\lambda_{1},-\lambda_{2})\\
T_{12}^{-} & =T_{1}(x_{1},0|-\lambda_{1})T_{2}(x_{2},0|-\lambda_{2})\\
T_{12}^{+} & =T_{1}(0,x_{1}|\lambda_{1})T_{2}(0,x_{2}|\lambda_{2})
\end{align*}
The existence of infinitely many conserved charges in involution requires
that the following expression has to vanish. 
\begin{multline}
\bigl[r_{12}(0|\lambda_{1},\lambda_{2}),\kappa_{1}(\lambda_{1})\kappa_{2}(\lambda_{2})\bigr]+\kappa_{1}(\lambda_{1})r_{12}(0|\lambda_{1},-\lambda_{2})\kappa_{2}(\lambda_{2})-\kappa_{2}(\lambda_{2})r_{12}(0|\lambda_{1},-\lambda_{2})\kappa_{1}(\lambda_{1})+\\
+\frac{1}{2}\Bigl(G_{12}(-\lambda_{1},\lambda_{2})\kappa_{1}(\lambda_{1})-\kappa_{1}(\lambda_{1})G_{12}(\lambda_{1},\lambda_{2})-G_{21}(-\lambda_{2},\lambda_{1})\kappa_{2}(\lambda_{2})+\kappa_{2}(\lambda_{2})G_{21}(\lambda_{2},\lambda_{1})\Bigr)=0\label{cbYBE}
\end{multline}
This is the \emph{classical boundary Yang-Baxter equation} (cbYBE).
If the $\kappa$-matrix fulfill this equation then the Poisson-bracket
of the double row monodromy matrix is 
\begin{equation}
\{\Omega_{1}(x_{1}|\lambda_{1}),\Omega_{2}(x_{2}|\lambda_{2})\}=t_{12}^{-}\left(\left[R_{12},\omega_{12}\right]+\omega_{1}^{(1)}\tilde{R}_{12}\omega_{2}^{(2)}-\omega_{2}^{(2)}\tilde{R}_{12}\omega_{1}^{(2)}\right)t_{12}^{+}
\end{equation}
This Poisson-bracket satisfies the Jacobi identity (this can be derived
by a straightforward but very long calculation). Using the split-point
procedure and the symmetric limit we can calculate the ``weak''
Poisson algebra of the global double row monodromy matrix \eqref{eq:globO}.
\begin{multline}
\{\Omega_{1}(\lambda_{1}),\Omega_{2}(\lambda_{2})\}=\left[r_{12}(-\infty|\lambda_{1},\lambda_{2}),\Omega_{1}(\lambda_{1})\Omega_{2}(\lambda_{2})\right]+\\
+\Omega_{1}(\lambda_{1})r_{12}(-\infty|\lambda_{1},-\lambda_{2})\Omega_{1}(\lambda_{1})-\Omega_{2}(\lambda_{2})r_{12}(-\infty|\lambda_{1},-\lambda_{2})\Omega_{1}(\lambda_{1})\label{eq:PoissonOm}
\end{multline}
Taking trace we get
\[
\{\Tr[\Omega(\lambda_{1})],\Tr[\Omega(\lambda_{2})]\}=0,
\]
which means we have infinite many conserved charges in involution.

\subsection{Poisson bracket in PCMs}

Let us specify now the previous findings for the PCMs. The Poisson-algebra
of the currents is the following \citep{Faddeev:1987ph,Maillet:1985ec}:
\begin{align}
\{J_{0}(x)\overset{\otimes}{,}J_{0}(y)\} & =\bigl[C,J_{0}\otimes1\bigr]\delta(x-y),\nonumber \\
\{J_{0}(x)\overset{\otimes}{,}J_{1}(y)\} & =\bigl[C,J_{1}\otimes1\bigr]\delta(x-y)-C\delta'(x-y),\label{eq:currPoisson}\\
\{J_{1}(x)\overset{\otimes}{,}J_{1}(y)\} & =0\nonumber 
\end{align}
where $\Jc=\Jc^{A}T_{A}$ if $\left\{ T_{A}\right\} $ is a basis
in $\gn$ for which we can define an invariant bilinear form $\left\langle T_{A},T_{B}\right\rangle =-\frac{1}{2}\Tr[T_{A},T_{B}]=C_{AB}$
and $C=C^{AB}T_{A}\otimes T_{B}$ where $C^{AD}C_{DB}=\delta_{B}^{A}$.
This form can be used to define a totally anti-symmetric tensor from
the structure constant $f_{ABC}=C_{AD}f_{BC}^{D}$ where $[T_{A},T_{B}]=f_{AB}^{C}T_{C}$.
For semi-simple Lie-algebras there exists a basis for which $C_{AB}=\delta_{AB}$.
In this basis the structure constant is totally anti-symmetric $f_{ABC}=f_{BC}^{A}=f^{ABC}$
and the Poisson bracket looks like 
\begin{align*}
\{J_{0}^{A}(x),J_{0}^{B}(y)\} & =f^{ABC}J_{0}^{C}\delta(x-y),\\
\{J_{0}^{A}(x),J_{1}^{B}(y)\} & =f^{ABC}J_{1}^{C}\delta(x-y)-\delta^{AB}\delta'(x-y),\\
\{J_{1}^{A}(x),J_{1}^{B}(y)\} & =0
\end{align*}

In the following we will need the Poisson-bracket of the group element
$g$ and the current $J_{0}^{L/R}$. For this, we can use the following
formula
\[
g(x)=g(-\infty)\mathcal{P}\overrightarrow{\exp}\int_{-\infty}^{x}J_{1}^{R}(y)\dd y=g(-\infty)t(-\infty,x),
\]
where we used the definition: 
\[
t(x,y)=\mathcal{P}\overrightarrow{\exp}\int_{x}^{y}J_{1}^{R}(z)\dd z
\]
 and \eqref{eq:currPoisson}:
\begin{align*}
\{J_{0}^{R}(x)\overset{\otimes}{,}g(y)\} & =\left(1\otimes g(-\infty)\right)\int_{-\infty}^{y}\left(1\otimes t(-\infty,z)\right)\left\{ J_{0}^{R}(x)\overset{\otimes}{,}J_{1}^{R}(z)\right\} \left(1\otimes t(z,y)\right)\dd z=\\
 & =\left(1\otimes g(-\infty)\right)\int_{-\infty}^{y}\left(1\otimes t(-\infty,z)\right)\Bigl(-\left[C,1\otimes J_{1}^{R}(z)\right]\delta(x-z)+\\
 & \qquad\qquad\qquad\qquad\qquad\qquad\qquad\quad\quad+C\partial_{z}\delta(z-x)\Bigr)\left(1\otimes t(z,y)\right)\dd z=\\
 & =\left(1\otimes g(-\infty)\right)\int_{-\infty}^{y}\partial_{z}\left(\left(1\otimes t(-\infty,z)\right)\left(C\delta(z-x)\right)\left(1\otimes t(z,y)\right)\right)\dd z=\\
 & =\left(1\otimes g\right)C\delta(x-y).
\end{align*}
Therefore
\begin{align}
\left\{ J_{0}^{R}(x)\overset{\otimes}{,}g(y)\right\}  & =\left(1\otimes g\right)C\delta(x-y) & \left\{ J_{0}^{R}(x)\overset{\otimes}{,}g^{-1}(y)\right\} = & -C\left(1\otimes g^{-1}\right)\delta(x-y)\label{eq:commJg}\\
\left\{ J_{0}^{L}(x)\overset{\otimes}{,}g(y)\right\}  & =-C\left(1\otimes g\right)\delta(x-y) & \left\{ J_{0}^{L}(x)\overset{\otimes}{,}g^{-1}(y)\right\} = & \left(1\otimes g^{-1}\right)C\delta(x-y)\nonumber 
\end{align}

The Poisson brackets of the space-like component of the Lax operator
is \citep{Maillet:1985ec}: 
\begin{align*}
\{\mathcal{L}_{1}(x|\lambda_{1}),\mathcal{L}_{2}(y|\lambda_{2})\}=- & \bigl[r_{12}(\lambda_{1},\lambda_{2}),\mathcal{L}_{1}(\lambda_{1})+\mathcal{L}_{2}(\lambda_{2})\bigr]\delta(x-y)+\\
+ & \bigl[s_{12}(\lambda_{1},\lambda_{2}),\mathcal{L}_{1}(\lambda_{1})-\mathcal{L}_{2}(\lambda_{2})\bigr]\delta(x-y)-\\
- & 2s(\lambda_{1},\lambda_{2})\delta'(x-y),
\end{align*}
where 
\begin{align*}
r(\lambda_{1},\lambda_{2}) & =-\frac{1}{2}\frac{1}{\lambda_{1}-\lambda_{2}}\frac{\lambda_{1}^{2}+\lambda_{2}^{2}-2\lambda_{1}^{2}\lambda_{2}^{2}}{(\lambda_{1}^{2}-1)(\lambda_{2}^{2}-1)}C,\\
s(\lambda_{1},\lambda_{2}) & =-\frac{1}{2}\frac{\lambda_{1}+\lambda_{2}}{(\lambda_{1}^{2}-1)(\lambda_{2}^{2}-1)}C.
\end{align*}
In \citep{Maillet:1985ec} a different convention is used which can
be obtained by the following changes: $\mathcal{L\to-\mathcal{L}},$$\lambda\to-\lambda$
, $\gamma\to-1$. This Poisson-bracket is the same as \eqref{eq:LPalg}
but in this special case the r- and s-matrices are space independent.

Furthermore, we can find a consistency check for the classical boundary
Yang-Baxter equation (cbYBE) in \ref{sec:Consistency} where
we prove that if $\kappa_{R}(\lambda)$ satisfies the cbYBE then $\kappa_{L}(\lambda)=g\kappa_{R}(1/\lambda)g^{-1}$
also does which has to follow from the inversion property of the reflection
matrices. In this derivation we have to use a non-trivial identity
of the $r$-matrix
\begin{equation}
r_{12}(\lambda_{1},\lambda_{2})=r_{12}(1/\lambda_{1},1/\lambda_{2})-\frac{1}{2}\left(\frac{\lambda_{1}}{1-\lambda_{1}^{2}}-\frac{\lambda_{2}}{1-\lambda_{2}^{2}}\right)C_{12}.\label{eq:idenr}
\end{equation}
In \ref{sec:Consistency} we also show that this identity
is a consequence of the inversion property and the $s$-matrix has
a similar property:
\[
s_{12}(\lambda_{1},\lambda_{2})=s_{12}(1/\lambda_{1},1/\lambda_{2})-\frac{1}{2}\left(\frac{\lambda_{1}}{1-\lambda_{1}^{2}}+\frac{\lambda_{2}}{1-\lambda_{2}^{2}}\right)C_{12}.
\]

In the following we solve the classical boundary Yang-Baxter equation
for constant $\kappa$-matrices.

\subsubsection{Constant $\kappa$-matrices}

Let $\kappa(\lambda)=U$ where $U\in G$ is a constant matrix. The
cbYBE can be written as
\[
\frac{1}{\lambda_{1}-\lambda_{2}}\left[C_{12},U_{1}U_{2}\right]+\frac{1}{\lambda_{1}+\lambda_{2}}\left(U_{1}C_{12}U_{2}-U_{2}C_{12}U_{1}\right)=0
\]
This equation has to be satisfied for every $\lambda_{1},\lambda_{2}\in\mathbb{C}$
therefore
\begin{align*}
\left[C_{12},U_{1}U_{2}\right] & =0 & \text{and} &  & U_{1}C_{12}U_{2} & =U_{2}C_{12}U_{1}
\end{align*}
The first equation is satisfied trivially because $C_{12}$ is invariant
i.e. $C_{12}=U_{1}U_{2}C_{12}U_{1}^{-1}U_{2}^{-2}$. Let us multiply
the second by $U_{1}$ from the left and by $U_{2}^{-1}$ from the
right 
\[
U_{1}^{2}C_{12}=U_{1}U_{2}C_{12}U_{1}U_{2}^{-1}=C_{12}U_{1}^{2}
\]
Using the explicit form of $C_{12}$ we obtain that
\begin{align*}
C^{AB}\left[X_{A},U^{2}\right]\otimes X_{B} & =0 & \Rightarrow &  & \left[X,U^{2}\right] & =0
\end{align*}
for all $X\in\mathfrak{g}$. Because we work with the defining representation
(which is irreducible), $U^{2}$ has to be proportional to the identity.
This is the same solution which we obtained from the analysis of the
boundary flatness equation. Therefore we can conclude that the consistent
solution of the flatness condition and the cbYBE are the same for
the constant $\kappa$-matrix.

In the end of the Subsection \ref{subsec:Lax-formalism-for}, we saw
that there is an other way to define a double row monodromy matrix:
\[
\Omega(\lambda)=T_{L}(-\lambda)^{-1}UT_{R}(\lambda).
\]
For this definition we should modify the formulas \eqref{cbYBE} and
\eqref{eq:PoissonOm}. However, this would require a long calculation.
Fortunately, we saw that there is another equivalent formalism of
this boundary condition:
\[
\Omega(\lambda)=T_{R}(-1/\lambda)\left(g^{-1}(0)U\right)T_{R}(\lambda)=T_{R}(-1/\lambda)\kappa(\lambda)T_{R}(\lambda).
\]

Using this, the generalization of \eqref{cbYBE} and \eqref{eq:PoissonOm}
are the following:
\begin{multline}
r_{12}(1/\lambda_{1},1/\lambda_{2})\kappa_{1}(\lambda_{1})\kappa_{2}(\lambda_{2})-\kappa_{1}(\lambda_{1})\kappa_{2}(\lambda_{2})r_{12}(\lambda_{1},\lambda_{2})+\\
+\kappa_{1}(\lambda_{1})r_{12}(\lambda_{1},-1/\lambda_{2})\kappa_{2}(\lambda_{2})+\kappa_{2}(\lambda_{2})r_{12}(-1/\lambda_{1},\lambda_{2})\kappa_{1}(\lambda_{1})+\\
+\frac{1}{2}\Bigl(G_{12}(-1/\lambda_{1},\lambda_{2})\kappa_{1}(\lambda_{1})-\kappa_{1}(\lambda_{1})G_{12}(\lambda_{1},\lambda_{2})-\\
-G_{21}(-1/\lambda_{2},\lambda_{1})\kappa_{2}(\lambda_{2})+\kappa_{2}(\lambda_{2})G_{21}(\lambda_{2},\lambda_{1})\Bigr)=0.\label{eq:cbYBE-2}
\end{multline}
\begin{multline}
\{\Omega_{1}(\lambda_{1}),\Omega_{2}(\lambda_{2})\}=r_{12}(1/\lambda_{1},1/\lambda_{2})\Omega_{1}(\lambda_{1})\Omega_{2}(\lambda_{2})-\Omega_{1}(\lambda_{1})\Omega_{2}(\lambda_{2})r_{12}(\lambda_{1},\lambda_{2})\\
+\Omega_{1}(\lambda_{1})r_{12}(1/\lambda_{1},-\lambda_{2})\Omega_{1}(\lambda_{1})-\Omega_{2}(\lambda_{2})r_{12}(\lambda_{1},-1/\lambda_{2})\Omega_{1}(\lambda_{1})\label{eq:PoissonOm-1}
\end{multline}

Let us check that the modified cbYBE \eqref{eq:cbYBE-2} is satisfied.
At first, let us calculate the $G$s.
\[
\left\{ \mathcal{L}_{1}(\lambda_{1}|x),\kappa_{2}(\lambda_{2})\right\} =-\frac{\lambda_{1}}{1-\lambda_{1}^{2}}\left\{ J_{0}(x)\overset{\otimes}{,}g^{-1}U\right\} =\frac{\lambda_{1}}{1-\lambda_{1}^{2}}C_{12}(g^{-1}U)_{2}\delta(x)
\]
therefore 
\[
G_{12}(\lambda_{1},\lambda_{2})=-\frac{\lambda_{1}}{1-\lambda_{1}^{2}}C_{12}(g^{-1}U)_{2}=-\frac{\lambda_{1}}{1-\lambda_{1}^{2}}C_{12}\kappa_{2}.
\]
Using this, the modified cbYBE \eqref{eq:cbYBE-2} looks like
\begin{multline}
r_{12}(1/\lambda_{1},1/\lambda_{2})\kappa_{1}\kappa_{2}-\kappa_{1}\kappa_{2}r_{12}(\lambda_{1},\lambda_{2})+\kappa_{1}r_{12}(\lambda_{1},-1/\lambda_{2})\kappa_{2}+\kappa_{2}r_{12}(-1/\lambda_{1},\lambda_{2})\kappa_{1}+\\
-\frac{1}{2}\left(\frac{\lambda_{1}}{1-\lambda_{1}^{2}}-\frac{\lambda_{2}}{1-\lambda_{2}^{2}}\right)C_{12}\kappa_{1}\kappa_{2}+\frac{1}{2}\frac{\lambda_{1}}{1-\lambda_{1}^{2}}\kappa_{1}C_{12}\kappa_{2}-\frac{1}{2}\frac{\lambda_{2}}{1-\lambda_{2}^{2}}\kappa_{2}C_{12}\kappa_{1}=0\label{eq:cbYBE-2-1}
\end{multline}
Using the identities \eqref{eq:idenr} and 
\[
r_{12}(\lambda_{1},-1/\lambda_{2})+\frac{1}{2}\frac{\lambda_{1}}{1-\lambda_{1}^{2}}=-\left(r_{12}(-1/\lambda_{1},\lambda_{2})-\frac{1}{2}\frac{\lambda_{2}}{1-\lambda_{2}^{2}}\right)=\tilde{r}_{12}(\lambda_{1},\lambda_{2}),
\]
the equation \eqref{eq:cbYBE-2-1} can be written as
\[
\left[r_{12}(\lambda_{1},\lambda_{2}),\kappa_{1}\kappa_{2}\right]+\kappa_{1}\tilde{r}_{12}(\lambda_{1},\lambda_{2})\kappa_{2}-\kappa_{2}\tilde{r}_{12}(\lambda_{1},\lambda_{2})\kappa_{1}=0
\]
where 
\[
\tilde{r}_{12}(\lambda_{1},\lambda_{2})=\frac{1}{2}\frac{\lambda_{1}\lambda_{2}-1}{\lambda_{1}\lambda_{2}+1}\frac{\lambda_{1}+\lambda_{2}}{(\lambda_{1}^{2}-1)(\lambda_{2}^{2}-1)}C_{12}.
\]
Therefore the modified cbYBE can be written as
\[
(\lambda_{1}\lambda_{2}+1)(2\lambda_{1}^{2}\lambda_{2}^{2}-\lambda_{1}^{2}-\lambda_{2}^{2})\left[C_{12}(\lambda_{1},\lambda_{2}),\kappa_{1}\kappa_{2}\right]-(\lambda_{1}-\lambda_{2})(\lambda_{1}+\lambda_{2})(\lambda_{1}\lambda_{2}-1)\left(\kappa_{1}C_{12}\kappa_{2}-\kappa_{2}C_{12}\kappa_{1}\right)=0
\]
Since the coefficients are linearly independent polynomials we have
\begin{align*}
\left[C_{12},\kappa_{1}\kappa_{2}\right] & =0 & \text{and} &  & \kappa_{1}C_{12}\kappa_{2} & =\kappa_{2}C_{12}\kappa_{1}.
\end{align*}
We have already solved these equations and the solution is $\kappa^{2}=e$
i.e. $g^{-1}Ug^{-1}U=e$ which is the same constraint what we get
from the boundary flatness equation. 

There is another consequence of the fact that we had to modify the
equation \eqref{eq:PoissonOm} to \eqref{eq:PoissonOm-1}. Now, the
traces of double row monodromy matrices are not in involution i.e.
\[
\left\{ \mathrm{Tr}[\Omega_{1}(\lambda_{1})],\mathrm{Tr}[\Omega_{1}(\lambda_{1})]]\right\} \neq0.
\]
Nevertheless one can show that there exists a conserved quantity $\mathcal{F}(\lambda)$
for which 
\[
\left\{ \mathcal{F}(\lambda_{1}),\mathcal{F}(\lambda_{2})]\right\} =0.
\]
 The explicit form being
\[
\mathcal{F}(\lambda)=\mathrm{Tr}\left[\Omega(1/\lambda)\Omega(\lambda)\right]=\mathrm{Tr}\left[T^{-1}(-\lambda)\kappa T(1/\lambda)T^{-1}(-1/\lambda)\kappa T(\lambda)\right]
\]

\subsubsection{Spectral parameter dependent $\kappa$-matrix}

The $\kappa$-matrices described in Section \ref{sec:Principal-Chiral-Models}
fulfill the classical boundary Yang-Baxter equation \eqref{cbYBE}.
The derivation can be found in \ref{sec:appBYBE}. 

In \citep{Gombor:2019bun} the following theorem was proven. 
\begin{thm*}
Let $U\in G$ for which $\mathrm{Ad}_{U}$ defines a Lie-algebra involution
and $\mathfrak{h}:=\left\{ X\in\mathfrak{g}|UXU^{-1}=X\right\} $.
If $\kappa(\lambda)$ is a solutions of the following cbYBE 
\[
\frac{1}{\lambda_{1}-\lambda_{2}}\left[C_{12},\kappa_{1}(\lambda)\kappa_{2}(\lambda)\right]+\frac{1}{\lambda_{1}+\lambda_{2}}\left(\kappa_{1}(\lambda)C_{12}\kappa_{2}(\lambda)-\kappa_{2}(\lambda)C_{12}\kappa_{1}(\lambda)\right)=0
\]
then $\kappa(\lambda)=U$ for semi-simple $\mathfrak{h}$ or $\kappa(\lambda)=U+\frac{1}{\lambda}X_{0}U+\mathcal{O}(\lambda^{-2})$
for reductive $\mathfrak{h}$ where $X_{0}$ is a central element
of $\mathfrak{h}$. The $\kappa$-matrix $\kappa(\lambda)$ is unique
for a given $U$ (up to normalization) if we fix the norm of $X_{0}$.
\end{thm*}
Previously we showed that these solutions exist therefore we classified
the field independent solutions of the cbYBE.

We close this subsection with the Poisson-algebra of the Noether charges
of the global symmetries. Let us start with the right charges
\begin{align*}
\tilde{Q}_{R}^{(0)} & =\Pi_{\hn}\left(Q_{R}^{(0)}\right)=\int_{-\infty}^{0}\Pi_{\hn}\left(J_{0}^{R}(x)\right)\dd x
\end{align*}
Using the Poisson-algebra of the current we can obtain that
\[
\left\{ \tilde{Q}_{R}^{(0)}\overset{\otimes}{,}\tilde{Q}_{R}^{(0)}\right\} =\left(\Pi_{\hn}\otimes\Pi_{\hn}\right)\circ\left[C,\tilde{Q}_{R}^{(0)}\otimes1\right],
\]
We can decompose the basis $\left\{ T_{A}\right\} $ into $\left\{ T_{a}\in\mathfrak{h}\right\} $
and $\left\{ T_{\alpha}\in\mathfrak{f}\right\} .$ Using these, the
equation above can be written as
\[
\left\{ \tilde{Q}_{R}^{(0)a}\overset{\otimes}{,}\tilde{Q}_{R}^{(0)b}\right\} =f^{abc}\tilde{Q}_{R}^{(0)c}
\]
therefore they form the Lie-algebra $\mathfrak{h}$ as expected. Let
us continue with the Noether charges of the left multiplication
\[
\tilde{Q}_{L}=Q_{L}-\frac{1}{2}(gMg^{-1})\Big|_{x=0}=\int_{-\infty}^{0}J_{0}^{L}(x)\dd x-\frac{1}{2}(gMg^{-1})\Big|_{x=0}
\]
The Poisson-bracket $\left\{ \tilde{Q}_{L}^{(0)}\overset{\otimes}{,}\tilde{Q}_{L}^{(0)}\right\} $
is not well defined because it contains the following expression
\[
\left\{ \int_{-\infty}^{0}J_{0}^{L}(x)\dd x\overset{\otimes}{,}\left(gMg^{-1}\right)\Big|_{x=0}\right\} 
\]
therefore we have to use the symmetric limit ($\Delta<0$):
\begin{multline*}
\left\{ \int_{-\infty}^{0}J_{0}^{L}(x)\dd x\overset{\otimes}{,}\left(gMg^{-1}\right)\Big|_{x=0}\right\} \coloneqq\\
\frac{1}{2}\lim_{\Delta\to0}\left(\left\{ \int_{-\infty}^{0}J_{0}^{L}(x)\dd x\overset{\otimes}{,}\left(gMg^{-1}\right)\Big|_{x=\Delta}\right\} +\left\{ \int_{-\infty}^{\Delta}J_{0}^{L}(x)\dd x\overset{\otimes}{,}\left(gMg^{-1}\right)\Big|_{x=0}\right\} \right)=\\
=\frac{1}{2}\lim_{\Delta\to0}\left\{ \int_{-\infty}^{0}J_{0}^{L}(x)\dd x\overset{\otimes}{,}\left(gMg^{-1}\right)\Big|_{x=\Delta}\right\} =\frac{1}{2}\left[C,\left(gMg^{-1}\right)\Big|_{x=0}\otimes1\right]
\end{multline*}
Using this, we can obtain the following equation
\[
\left\{ \tilde{Q}_{L}^{(0)}\overset{\otimes}{,}\tilde{Q}_{L}^{(0)}\right\} =\left[C,\tilde{Q}_{L}^{(0)}\otimes1\right],
\]
which can be written as
\[
\left\{ \tilde{Q}_{L}^{(0)A}\overset{\otimes}{,}\tilde{Q}_{L}^{(0)B}\right\} =f^{ABC}\tilde{Q}_{L}^{(0)C}.
\]
Clearly these charges form the Lie-algebra $\mathfrak{g}$ as expected.
This calculation shows the importance of the symmetric limit because
if we do not use it properly then we cannot get the proper Poisson-algebra
of the Noether charges of the symmetry $G_{L}$. 

\subsection{Poisson bracket in $\mathrm{O}(N)$ sigma models\label{subsec:Poisson-bracket-in}}

The Poisson-algebra of the fields $n_{i}$ is the following
\begin{align*}
\left\{ n_{i}(x),n_{j}(y)\right\}  & =0\\
\left\{ \dot{n}_{i}(x),n_{j}(y)\right\}  & =\left(\delta_{ij}-n_{i}n_{j}\right)\delta(x-y)\\
\left\{ \dot{n}_{i}(x),\dot{n}_{j}(y)\right\}  & =\left(n_{i}\dot{n}_{j}-\dot{n}_{i}n_{j}\right)\delta(x-y)
\end{align*}
From this one can calculate the Poisson-algebra of the currents \citep{Maillet:1985fn}:
\begin{align*}
\left\{ \hat{J}_{0}(x)\overset{\otimes}{,}\hat{J}_{0}(y)\right\}  & =\left[C,\hat{J}_{0}(x)\otimes1\right]\delta(x-y)\\
\left\{ \hat{J}_{0}(x)\overset{\otimes}{,}\hat{J}_{1}(y)\right\}  & =\left[C,\hat{J}_{1}(x)\otimes1\right]\delta(x-y)-2\Gamma(y)\delta'(x-y)\\
\left\{ \hat{J}_{1}(x)\overset{\otimes}{,}\hat{J}_{1}(y)\right\}  & =0
\end{align*}
where 
\begin{align*}
C & =2(K-P)\\
\Gamma(x) & =C\left(Z(x)\otimes1\right)+\left(Z(x)\otimes1\right)C=C\left(1\otimes Z(x)\right)+\left(1\otimes Z(x)\right)C
\end{align*}
and $(P)_{ij,kl}=\delta_{il}\delta_{jk}$, $(K)_{ij,kl}=\delta_{ik}\delta_{jl}$
are the permutation and the trace operators and $(Z)_{ij}=n_{i}n_{j}$.

Using this, one can obtain the non-ultralocal Poisson-algebra of the
space-like component of the Lax-connection \eqref{eq:LPalg} where
the r- and s-matrices are 
\begin{align*}
r(x|\lambda_{1},\lambda_{2}) & =\frac{\lambda_{1}\lambda_{2}}{(\lambda_{1}-\lambda_{2})(\lambda_{1}\lambda_{2}-1)}C+\frac{(\lambda_{1}-\lambda_{2})}{(\lambda_{1}^{2}-1)(\lambda_{2}^{2}-1)}\frac{(\lambda_{1}\lambda_{2}+1)}{(\lambda_{1}\lambda_{2}-1)}\Gamma(x)\\
s(x|\lambda_{1},\lambda_{2}) & =\frac{(\lambda_{1}+\lambda_{2})}{(\lambda_{1}^{2}-1)(\lambda_{2}^{2}-1)}\Gamma(x)
\end{align*}
At first, we solve the cbYBE for constant $\kappa$-matrices and after
that we check the spectral parameter and field dependent $\kappa$-matrix.

\subsubsection{Constant $\kappa$-matrix}

For $\kappa(\lambda)=U\in\mathrm{O}(N)$, the cbYBE looks like
\[
\left[r_{12}(\lambda_{1},\lambda_{2}),U_{1}U_{2}\right]+U_{1}r_{12}(\lambda_{1},-\lambda_{2})U_{2}-U_{2}r_{12}(\lambda_{1},-\lambda_{2})U_{1}=0.
\]
After substitution, we obtain the following four equations:
\begin{align*}
\left[C_{12},U_{1}U_{2}\right] & =0\\
U_{1}C_{12}U_{2} & =U_{2}C_{12}U_{1}\\
\left[\Gamma_{12},U_{1}U_{2}\right] & =0\\
U_{1}\Gamma_{12}U_{2} & =U_{2}\Gamma_{12}U_{1}
\end{align*}
The first equation follows from the fact that $U\in\mathrm{O}(N)$.
From the second equation if follows that $U^{2}=\pm1$ i.e. $U=\pm U^{T}$.
Multiplying the fourth one by $U_{1}$ from the left and right, we
can see that the third one comes from the fourth. Let us write the
third one explicitly.
\begin{align*}
C_{12}Z_{2}U_{1}U_{2}+Z_{2}C_{12}U_{1}U_{2} & =U_{1}U_{2}C_{12}Z_{2}+U_{1}U_{2}Z_{2}C_{12}
\end{align*}
Multiplying by $U_{1}^{T}U_{2}^{T}$from the left, we obtain the following
\[
C_{12}U_{2}^{T}Z_{2}U_{1}+U_{2}^{T}Z_{2}U_{2}C_{12}=C_{12}Z_{2}+Z_{2}C_{12}.
\]
Using the explicit form of $C_{12}$, we can obtain that
\[
(P_{12}-K_{12})(Z_{2}-U_{2}^{T}Z_{2}U_{2})=(Z_{2}-U_{2}^{T}Z_{2}U_{2})(K_{12}-P_{12}).
\]
Let us multiply by $P_{12}$ from the left.
\[
\tilde{Z}_{2}-K_{12}\tilde{Z}_{2}=K_{12}\tilde{Z}_{2}-\tilde{Z}_{1}
\]
where $\tilde{Z}=Z-U^{T}ZU$. Taking the trace on the first site:
\[
N\tilde{Z}-\tilde{Z}=\tilde{Z}^{T}-\mathrm{Tr}\left(\tilde{Z}\right)1
\]
Using that $\tilde{Z}_{2}^{T}=\tilde{Z}_{2}$, $\mathrm{Tr}\left(\tilde{Z}\right)=0$
and $N>2$, we obtain that
\[
\tilde{Z}=0.
\]
Since $U$ can be $U=\pm U^{T}$, there are two cases.
\begin{enumerate}
\item $U=U^{T}$. Using a global symmetry transformation $U$ can be diagonalized
as
\[
U=\left(\begin{array}{cc}
1_{N-k} & 0_{N-k\times k}\\
0_{k\times k} & -1_{k}
\end{array}\right)
\]
and $Z$ in the same block diagonal form looks like
\[
Z=\left(\begin{array}{cc}
\nt\nt^{T} & \nt\nh^{T}\\
\nh\nt^{T} & \nh\nh^{T}
\end{array}\right)
\]
therefore $\tilde{Z}$ looks like
\[
\tilde{Z}=\left(\begin{array}{cc}
0 & 2\nt\nh^{T}\\
2\nh\nt^{T} & 0
\end{array}\right).
\]
From this explicit form we can see that $\tilde{Z}=0$ if and only
if $\nt=0$ or $\nh=0$. 
\item $U=-U^{T}$. Using a global symmetry transformation $U$ can be diagonalized
as
\[
U=\left(\begin{array}{cc}
0_{n\times n} & -1_{n}\\
1_{n} & 0_{n\times n}
\end{array}\right)
\]
where $n=N/2$ and $\tilde{Z}$ looks like
\[
\tilde{Z}=\left(\begin{array}{cc}
\nt\nt^{T}-\nh\nh^{T} & \nt\nh^{T}+\nh\nt^{T}\\
\nt\nh^{T}+\nh\nt^{T} & \nh\nh^{T}-\nt\nt^{T}
\end{array}\right).
\]
Multiplying the off-diagonal terms by $\nh$ form the right, we obtain
\[
\nt\left(\nh^{T}\nh\right)=-\nh\left(\nt^{T}\nh\right)
\]
and multiplying this by $\nh^{T}$ form the left, we obtain 
\[
\left(\nh^{T}\nh\right)\left(\nt^{T}\nh\right)=0.
\]
At first, let us assume that $\nh\neq0$ therefore $\nt^{T}\nh=0$.
Substituting this to the previous equation, we obtain that $\nt=0$.
Using this in the diagonal term, we obtain that $\nh\nh^{T}=0$ which
contradicts to $\nh\neq0$. Therefore $\nh=0.$ From $\n^{T}\n=1$
and from the diagonal, we obtain that $\nt^{T}\nt=1$ and $\nt\nt^{T}$
which is a contradiction. Therefore anti-symmetric $U$ cannot be
a solution of the cbYBE.
\end{enumerate}
We can conclude that we have obtained the same constant $\kappa$-matrices
from the cbYBE as we got from the boundary flatness condition. 

\subsubsection{Spectral parameter and field dependent $\kappa$-matrix}

If we want to check that the new $\kappa$-matrix \eqref{kappaOn}
satisfy the classical boundary Yang-Baxter equation \eqref{cbYBE}
then we have to compute $G_{12}(\lambda_{1},\lambda_{2})$. For this,
we will need the following Poisson brackets:
\begin{align*}
\left\{ \left(J_{0}(x)\right)_{ij},n_{k}(y)\right\}  & =2\left(\delta_{jk}n_{i}-\delta_{ik}n_{j}\right)\delta(x-y)\\
\left\{ J_{0}(x)\overset{\otimes}{,}h(y)\right\}  & =\left[1\otimes h,C\right]\delta(x-y)\\
\left\{ J_{0}(x)\overset{\otimes}{,}(hMh)(y)\right\}  & =\left(\left(1\otimes h\right)\left[C,1\otimes M\right]\left(1\otimes h\right)+\left[1\otimes hMh,C\right]\right)\delta(x-y)
\end{align*}
From this
\[
G(\lambda,\mu)\delta(x)=-\left\{ \mathcal{L}(x|\lambda)\overset{\otimes}{,}\kappa(\mu)\right\} =\frac{\lambda}{1-\lambda^{2}}\left(1\otimes\left(\frac{1}{\mu}+M\right)\right)\left\{ J_{0}(x)\overset{\otimes}{,}(hMh)(0)\right\} 
\]
therefore
\[
G_{12}(\lambda,\mu)=\frac{1}{\mu}\frac{\lambda}{1-\lambda^{2}}\left(1+\mu M_{2}\right)\left(h_{2}\left[C_{12},M_{2}\right]h_{2}+\left[\left(hMh\right)_{2},C_{12}\right]\right)
\]
We checked the cbYBE for $\mathrm{O}(4)$ and $\mathrm{O}(6)$ sigma
models with explicit calculations using Wolfram Mathematica. For this,
we parameterized the sphere with stereo-graphic coordinates:
\begin{align*}
n_{a} & =\frac{2\xi_{a}}{1+\xi^{2}}\qquad\text{for }a=1,\dots,N-1\\
n_{N} & =\frac{1-\xi^{2}}{1+\xi^{2}}
\end{align*}
where 
\[
\xi^{2}=\sum_{a=1}^{N-1}\xi_{a}\xi_{a}.
\]
Using this parameterization we can calculate explicitly the matrices
$r(\lambda,\mu),G(\lambda,\mu),\kappa(\lambda)$ and we can substitute
these into the cbYBE. Using Mathematica we have checked that the cbYBE
is satisfied for $\mathrm{O}(4)$ and $\mathrm{O}(6)$ sigma models.

\section{Conclusion}

In this paper new double row monodromy matrices have been determined
for the principal chiral models. The corresponding integrable boundary
conditions break one chiral half of the symmetry to $G_{L}\times H_{R}$
where $H_{R}$ was not arbitrary but $G/H_{R}$ had to be a symmetric
space and the Lie algebra of $H_{R}$ was not semi-simple. We determined
the boundary conditions which correspond to these monodromy matrices.
Both the monodromy matrices and boundary conditions contain free parameters.

We used these results for finding new monodromy matrices for the $\On{N}$
sigma models. At first, the $\SO{4}\cong\SU{2}_{L}\times\SU{2}_{R}$
isometry was used to determine the $\SU{2}_{L}\times\mathrm{U}(1)_{R}$
symmetric $\kappa$ matrices for $\SO{4}$ sigma models. These new
spectral parameter dependent $\kappa$ matrices were then generalized
for $\On{2n}$ sigma models. They corresponds to $\mathrm{U}(n)$
symmetric boundary conditions.

We also showed that these $\kappa$-matrices satisfy the classical
boundary Yang-Baxter equation therefore there exist infinitely many
conserved charges in involution i.e. the boundary conditions proportional
to these $\kappa$s are classically integrable.

There exist quantum $\On{4}$ sigma models which have reflection matrix
with two free parameters and the residual symmetry is $\On{2}\times\On{2}$
\citep{Arnaudon:2003gj}. Therefore one interesting direction to pursue
would be to find the classical field theoretical description of these
quantum theories i.e. $\kappa$ matrices and boundary conditions which
have two independent parameters and residual symmetry $\On{2}\times\On{2}$.
In the language of the $\SU{2}$ PCM, this means boundary conditions
which independently break left and right symmetries. These results
could be then generalized to general PCMs.

As a last remark, it would be interesting to check that the quantum
version of the $\kappa$ matrices determined in the paper are really
the known reflection matrices. This could be done in the large-N limit.
Recently, the large-N limit was studied for the $\mathbb{C}\mathrm{P}^{N}$
sigma models on finite intervals e.g. \citep{Bolognesi:2016zjp}\citep{Bolognesi:2018njt}.
These methods may also be applicable to the models studied in this
paper.

\addtocontents{toc}{\protect\setcounter{tocdepth}{1}}

\section*{Acknowledgment}

I thank Zoltán Bajnok and László Palla for the useful discussions
and for reading the manuscript. The work was supported by the NKFIH
116505 Grant.

\appendix

\section{Non-local conserved charges\label{subsec:Non-local-conserved-charges}}

If we expand the monodromy matrix around $\lambda=\lambda_{0}$ we
get infinitely many conserved charges which are generally non-local.
In this section we will deal with the expansions around $\lambda=\infty$
and $\lambda=0$ and we will give the first two terms of these series.

\subsection{Expansion around $\lambda=\infty$}

We will start with the expansion of the one row monodromy matrix 
\begin{multline}
T_{R}(\lambda)=\Pexp\left(\int_{-\infty}^{0}-\mathcal{L}^{R}(\lambda)\dd x\right)=\mathrm{exp}\left(\sum_{r=0}^{\infty}\left(-\frac{1}{\lambda}\right)^{r+1}Q_{R}^{(r)}\right)=\\
1-\frac{1}{\lambda}Q_{R}^{(0)}+\frac{1}{\lambda^{2}}\left(Q_{R}^{(1)}+\frac{1}{2}Q_{R}^{(0)2}\right)+\dots\label{seriesT}
\end{multline}
Since
\[
\mathcal{L}^{R}(\lambda)=\frac{1}{1-\lambda^{2}}J_{1}^{R}-\frac{\lambda}{1-\lambda^{2}}J_{0}^{R}=\frac{1}{\lambda}J_{0}^{R}-\frac{1}{\lambda^{2}}J_{1}^{R}+\dots
\]
the expansion leads to
\[
T_{R}(\lambda)=1-\frac{1}{\lambda}\int_{-\infty}^{0}J_{0}^{R}(x)\dd x+\frac{1}{\lambda^{2}}\left(\int_{-\infty}^{0}J_{1}^{R}(x)\dd x+\int_{-\infty}^{0}\int_{-\infty}^{x_{1}}J_{0}^{R}(x_{1})J_{0}^{R}(x_{2})\dd x_{1}\dd x_{2}\right)+\dots
\]
which gives the first two charges
\begin{align*}
Q_{R}^{(0)} & =\int_{-\infty}^{0}J_{0}^{R}(x)\dd x,\\
Q_{R}^{(1)} & =\int_{-\infty}^{0}J_{1}^{R}(x)\dd x+\frac{1}{2}\int_{-\infty}^{0}\int_{-\infty}^{x_{1}}\left[J_{0}^{R}(x_{1}),J_{0}^{R}(x_{2})\right]\dd x_{1}\dd x_{2}.
\end{align*}
In order to calculate the expansion of the monodromy matrix we will
also need the following series: 
\begin{equation}
T_{R}^{-1}(-\lambda)=\mathrm{exp}\left(-\sum_{r_{0}}^{\infty}\left(\frac{1}{\lambda}\right)^{r+1}Q_{R}^{(r)}\right)=1-\frac{1}{\lambda}Q_{R}^{(0)}+\frac{1}{\lambda^{2}}\left(-Q_{R}^{(1)}+\frac{1}{2}Q_{R}^{(0)2}\right)+\dots,\label{seriesTi}
\end{equation}

In Subsection \ref{subsec:Spectral-parameter-dependent}, the classification
of the new $\kappa$-matrices are showed. For $\gn=\so{2n},\spp{n}$,
$\hn=\ue\oplus\su{n}$ the form of these are the same: 
\[
\kappa(\lambda)=\frac{1}{\sqrt{1+a^{2}\lambda^{2}}}\left(1+\lambda M\right),
\]
where $M$ generates the $\ue$ and $M^{2}=-a^{2}1$ and $M=-M^{T}$.
The generalization for other $\kappa$ matrices follows straightforwardly.
The expansion of the $\kappa$ is the following: 
\begin{equation}
\kappa(\lambda)=U+\frac{1}{a\lambda}-\frac{1}{2}\frac{1}{(a\lambda)^{2}}U+\dots\label{seriesK}
\end{equation}
where $M=aU$. The conserved charges come from the expansion of the
double row monodromy matrix. 
\begin{multline*}
\Omega(\lambda)=T_{R}^{-1}(-\lambda)\kappa(\lambda)T_{R}(\lambda)=U\cdot\mathrm{exp}\left(2\sum_{r_{0}}^{\infty}\left(-\frac{1}{\lambda}\right)^{r+1}\tilde{Q}_{R}^{(r)}\right)=\\
U-\frac{2}{\lambda}U\tilde{Q}_{R}^{(0)}+\frac{2}{\lambda^{2}}\left(U\tilde{Q}_{R}^{(1)}+U\tilde{Q}_{R}^{(0)2}\right)+\dots,
\end{multline*}
where $\{\tilde{Q}_{R}^{(r)}\}$ is the infinite set of conserved
charges. In the above equation multiplication with $U$ is necessary
for the proper normalization because 
\[
\lim_{\lambda\to\infty}\Omega(\lambda)=U.
\]
Using \eqref{seriesT}, \eqref{seriesTi} and \eqref{seriesK}: 
\begin{multline*}
\Omega(\lambda)=\left[1-\frac{1}{\lambda}Q_{R}^{(0)}+\frac{1}{\lambda^{2}}\left(-Q_{R}^{(1)}+\frac{1}{2}Q_{R}^{(0)2}\right)+\dots\right]\cdot\\
\left[U+\frac{1}{a\lambda}-\frac{1}{2}\frac{1}{(a\lambda)^{2}}U+\dots\right]\cdot\left[1-\frac{1}{\lambda}Q_{R}^{(0)}+\frac{1}{\lambda^{2}}\left(Q_{R}^{(1)}+\frac{1}{2}Q_{R}^{(0)2}\right)+\dots,\right]=\\
=U-\frac{1}{\lambda}U\left(Q_{R}^{(0)}+U^{T}Q_{R}^{(0)}U-\frac{1}{a}U^{T}\right)+\\
+\frac{1}{\lambda^{2}}U\left(Q_{R}^{(1)}-U^{T}Q_{R}^{(1)}U+\frac{1}{2}(Q_{R}^{(0)2}+U^{T}Q_{R}^{(0)2}U)+U^{T}Q_{R}^{(0)}UQ_{R}^{(0)}-\frac{2}{a}U^{T}Q_{R}^{(0)}-\frac{1}{2a^{2}}\right)
\end{multline*}
From this the first two conserved charges are the following: 
\begin{align*}
\tilde{Q}_{R}^{(0)} & =\Pi_{\hn}\left(Q_{R}^{(0)}\right)+\frac{1}{2a}U,\\
\tilde{Q}_{R}^{(1)} & =\Pi_{\fn}\left(Q_{R}^{(1)}\right)+\frac{1}{2}\left[\Pi_{\hn}\left(Q_{R}^{(0)}\right),\Pi_{\fn}\left(Q_{R}^{(0)}\right)\right]+\frac{1}{2a}\left[U,Q_{R}^{(0)}\right]=\\
 & =\Pi_{\fn}\left(Q_{R}^{(1)}\right)+\frac{1}{2}\left[\Pi_{\hn}\left(Q_{R}^{(0)}\right)+\frac{1}{a}U,\Pi_{\fn}\left(Q_{R}^{(0)}\right)\right].
\end{align*}
The first charge is equivalent to the charge \eqref{QR} (up to a
constant). $\tilde{Q}_{R}^{(1)}$ is very similar to the charge for
the $g\in H$ restricted boundary condition but there is an extra
term: $\left[U,Q_{R}^{(0)}\right]$ \citep{MacKay:2004tc}. These
charges also satisfy the relations: $\tilde{Q}_{R}^{(0)}\in\mathfrak{h}$
and $\tilde{Q}_{R}^{(1)}\in\mathfrak{f}$.

For a crosscheck we can take the time derivative of these charges
and we will see that they all vanish.

\subsection{Expansion around $\lambda=0$}

For the expansion around $\lambda=0$, we can use the inversion property
of the double row monodromy matrix \eqref{eq:invOmega}: 
\begin{multline}
\Omega_{R}(\lambda)=g^{-1}(-\infty)\left(T_{L}^{-1}(-1/\lambda)\left(g\kappa(\lambda)g^{-1}\right)\Big|_{x=0}T_{L}(1/\lambda)\right)g(-\infty)=\\
g^{-1}(-\infty)\mathrm{exp}\left(2\sum_{r_{0}}^{\infty}(-\lambda)^{r+1}\tilde{Q}_{L}^{(r)}\right)g(-\infty)\label{eq:LR}
\end{multline}
We can do the same calculation as before: 
\begin{multline*}
T_{L}^{-1}(-1/\lambda)\left(g\kappa(\lambda)g^{-1}\right)\Big|_{x=0}T_{L}(1/\lambda)=\\
=1-2\lambda\left(Q_{L}^{(0)}-\frac{1}{2}gMg^{-1}\Big|_{x=0}\right)+2\lambda^{2}\left(Q_{L}^{(0)2}-\frac{1}{2}\left[Q_{L}^{(0)},gMg^{-1}\Big|_{x=0}\right]_{+}-\frac{1}{4}a^{2}\right)+\dots
\end{multline*}
therefore the conserved charges are the following: 
\begin{align*}
\tilde{Q}_{L}^{(0)} & =Q_{L}^{(0)}-\frac{1}{2}gMg^{-1}\Big|_{x=0},\\
\tilde{Q}_{L}^{(1)} & =0.
\end{align*}
We can see that the first conserved charge is equal to the Noether
charge of the left multiplication symmetry \eqref{QL}: $\tilde{Q}_{L}^{(0)}=\tilde{Q}_{L}$.
The second set of charges vanish. This is similar to the case of the
free boundary condition ($\gn=\hn$) in \citep{MacKay:2004tc}.

\section[cbYBE for the new $\kappa$s]{Classical boundary Yang-Baxter equation for the new $\kappa$s\label{sec:appBYBE}}

In this section, we prove that matrices described in Subsection \ref{subsec:Spectral-parameter-dependent}
fulfill the cbYBE \eqref{cbYBE}.

We start with the $N=0$ case. For this, the cbYBE \eqref{cbYBE}
looks like: 
\begin{multline}
\frac{1}{\lambda_{1}-\lambda_{2}}\bigl[C_{12},(1+\lambda_{1}M_{1})(1+\lambda_{2}M_{2})\bigr]+\\
+\frac{1}{\lambda_{1}+\lambda_{2}}\biggl((1+\lambda_{1}M_{1})C_{12}(1+\lambda_{2}M_{2})-(1+\lambda_{2}M_{2})C_{12}(1+\lambda_{1}M_{1})\biggr)\overset{?}{=}0.\label{cbYBE2}
\end{multline}
This equation is satisfied thanks to the following identities: 
\begin{align}
[C_{12},M_{1}] & =-[C_{12},M_{2}],\label{ii1}\\{}
[C_{12},M_{1}M_{2}] & =0\label{ii2}\\
M_{1}C_{12}M_{2} & =M_{2}C_{12}M_{1}.\label{ii3}
\end{align}
Equation \eqref{ii1} follows from $M\in\gn$. 
\[
[C_{12},M_{1}]=[T_{A},M^{B}T_{B}]\otimes T^{A}=f_{AB}^{C}M^{B}T_{C}\otimes T^{A}=-M^{B}T_{C}\otimes[T^{C},T_{B}]=-[C_{12},M_{2}].
\]
Equation \eqref{ii2} and \eqref{ii3} follows from $M^{2}\sim1$
which means $MT_{a}M^{-1}=T_{a}$ and $MT_{\alpha}M^{-1}=-T_{\alpha}$
where $T_{a}\in\hn$ and $T_{\alpha}\in\fn$. 
\begin{multline*}
[C_{12},M_{1}M_{2}]=T_{A}M\otimes T^{A}M-MT_{A}\otimes MT^{A}=\\
=T_{A}M\otimes T^{A}M-(T_{a}M\otimes T^{a}M+(-T_{\alpha}M)\otimes(-T^{\alpha}M)=0.
\end{multline*}
The derivation of \eqref{ii3} is similar.

In the following we will continue with the $N\neq0$ case. The cbYBE
looks like: 
\begin{multline}
\frac{1}{\lambda_{1}-\lambda_{2}}\bigl[C_{12},(1+\lambda_{1}M_{1}+\lambda_{1}^{2}N_{1})(1+\lambda_{2}M_{2}+\lambda_{2}^{2}N_{2})\bigr]+\\
+\frac{1}{\lambda_{1}+\lambda_{2}}\biggl((1+\lambda_{1}M_{1}+\lambda_{1}^{2}N_{1})C_{12}(1+\lambda_{2}M_{2}+\lambda_{2}^{2}N_{2})-\\
-(1+\lambda_{2}M_{2}+\lambda_{2}^{2}N_{2})C_{12}(1+\lambda_{1}M_{1}+\lambda_{1}^{2}N_{1})\biggr)\overset{?}{=}0.\label{cbYBE3}
\end{multline}
The matrices $M$ and $N$ satisfy the following identities: 
\begin{align}
[C_{12},M_{1}] & =-[C_{12},M_{2}],\label{id1}\\{}
[C_{12},M_{1}M_{2}] & =-[C_{12},N_{1}]-[C_{12},N_{2}],\label{id2}\\
M_{1}C_{12}M_{2}-M_{2}C_{12}M_{1} & =-[C_{12},N_{1}]+[C_{12},N_{2}],\label{id3}\\{}
[C_{12},M_{1}N_{2}] & =-[C_{12},N_{1}M_{2}],\label{id4}\\{}
[C_{12},M_{1}N_{2}] & =M_{1}C_{12}N_{2}-N_{2}C_{12}M_{1}.\label{id5}\\
\left[C_{12},N_{1}N_{2}\right] & =0,\label{id6}\\
N_{1}C_{12}N_{2} & =N_{2}C_{12}N_{1}.\label{id7}
\end{align}
Using these, the equation \eqref{cbYBE3} is satisfied. The identity
\eqref{id1} is satisfied because $M\in\gn$. Let us see \eqref{id2}
and \eqref{id3}. 
\begin{multline*}
[C_{12},M_{1}M_{2}]-M_{1}C_{12}M_{2}+M_{2}C_{12}M_{1}=\left[[C_{12},M_{1}],M_{2}\right]_{+}=\\
=-\left[[C_{12},M_{2}],M_{2}\right]_{+}=-[C_{12},M_{2}^{2}]=-2[C_{12},N_{2}]
\end{multline*}
\begin{multline*}
[C_{12},M_{1}M_{2}]+M_{1}C_{12}M_{2}-M_{2}C_{12}M_{1}=\left[[C_{12},M_{2}],M_{1}\right]_{+}=\\
=-\left[[C_{12},M_{1}],M_{1}\right]_{+}=-[C_{12},M_{1}^{2}]=-2[C_{12},N_{1}]
\end{multline*}
where we used \eqref{condMN}. By adding and subtracting the equations
above we can get \eqref{id2} and \eqref{id3}. Equations \eqref{id6}
and \eqref{id7} follows from $N^{2}\sim1$ similarly to \eqref{ii2}
and \eqref{ii3}.

Now we only have to prove the equation \eqref{id4} and \eqref{id5}.
This can be done by using the explicit forms of $M$ and $N$ which
were shown in Subsection \ref{subsec:Spectral-parameter-dependent}.
When $N\neq0$, $(MN-c1)\in\gn$ where $c$ is a number. For ($\gn=\su{n}$,
$\hn=\ue\oplus\su{m}\oplus\su{n-m}$) $c=i\frac{4km}{\lambda n}$
and for ($\gn=\so{n}$, $\hn=\so{2}\oplus\so{n-2}$) $c=0$. 
\begin{multline}
[C_{12},M_{1}N_{2}]=(C_{12}M_{1}N_{1}^{-1}-M_{1}N_{2}C_{12}N_{1}^{-1}N_{2}^{-1})N_{1}N_{2}=\\
=(C_{12}M_{1}N_{1}^{-1}-M_{1}N_{1}^{-1}C_{12})N_{1}N_{2}=[C_{12},M_{1}N_{1}^{-1}]N_{1}N_{2}=-[C_{12},M_{2}N_{2}^{-1}]N_{1}N_{2}=\\
=-(C_{12}N_{1}M_{2}-M_{2}N_{2}^{-1}C_{12}N_{1}N_{2})=-(C_{12}N_{1}M_{2}-M_{2}N_{1}C_{12})=-[C_{12},N_{1}M_{2}],
\end{multline}
where we used $MN-c1\in\gn$ and \eqref{id7}. Finally let us see
the derivation of \eqref{id5}: 
\begin{multline*}
[C_{12},M_{1}N_{2}]=\left[X_{a},M\right]\otimes X^{a}N+\left[X_{\alpha},M\right]_{+}\otimes X^{\alpha}N=\\
=\left[M,X_{a}\right]\otimes X^{a}N+\left[X_{\alpha},M\right]_{+}\otimes X^{\alpha}N=M_{1}C_{12}N_{2}-N_{2}C_{12}M_{1}
\end{multline*}
where we used that $\left[M,X_{a}\right]=0$ for all $X_{a}\in\mathfrak{h}$.

\section{Consistency check of the cbYBE\label{sec:Consistency}}

In the PCM we can work with right or left currents. For a general
boundary condition the $\kappa$-matrices can be different using right
or left currents. Let $\kappa^{L}$ and $\kappa^{R}$ be these two
$\kappa$-matrices. We saw that the double row monodromy matrices
and the $\kappa$-matrices have the inversion property:
\begin{align*}
\Omega_{L}(\lambda) & =g(-\infty)\Omega_{R}(1/\lambda)g^{-1}(-\infty),\\
\kappa^{L}(\lambda) & =g(0)\kappa^{R}(1/\lambda)g^{-1}(0).
\end{align*}
The classical boundary Yang-Baxter equation (cbYBE) for $\kappa^{R}(\lambda)$
and $\kappa^{L}(\lambda)$ are the following:
\begin{multline}
\bigl[r_{12}(\lambda_{1},\lambda_{2}),\kappa_{1}^{L/R}(\lambda_{1})\kappa_{2}^{L/R}(\lambda_{2})\bigr]+\\
+\kappa_{1}^{L/R}(\lambda_{1})r_{12}(\lambda_{1},-\lambda_{2})\kappa_{2}^{L/R}(\lambda_{2})-\kappa_{2}^{L/R}(\lambda_{2})r_{12}(\lambda_{1},-\lambda_{2})\kappa_{1}^{L/R}(\lambda_{1})+\\
+\frac{1}{2}\Bigl(G_{12}^{L/R}(-\lambda_{1},\lambda_{2})\kappa_{1}^{L/R}(\lambda_{1})-\kappa_{1}^{L/R}(\lambda_{1})G_{12}^{L/R}(\lambda_{1},\lambda_{2})-\\
-G_{21}^{L/R}(-\lambda_{2},\lambda_{1})\kappa_{2}^{L/R}(\lambda_{2})+\kappa_{2}^{L/R}(\lambda_{2})G_{21}^{L/R}(\lambda_{2},\lambda_{1})\Bigr)=0\label{cbYBE-1}
\end{multline}
where we assumed that
\[
\left\{ \mathcal{L}_{1}^{L/R}(x|\lambda_{1}),\kappa_{2}^{L/R}(\lambda_{2})\right\} =-G_{12}^{L/R}(\lambda_{1},\lambda_{2})\delta(x).
\]
This assumption implicitly contains that $\kappa^{L/R}$ does not
depend on the time derivative of the fields.

In the following we prove that if $\kappa^{L}(\lambda)$ satisfies
the cbYBE then $\kappa^{R}(\lambda)=g^{-1}\kappa^{L}(1/\lambda)g$
also does. At first let us assume that $\kappa^{L}(\lambda)$ satisfies
the cbYBE \eqref{cbYBE-1}. Let us see $G_{12}^{L}$:
\begin{multline}
G_{12}^{L}(\lambda_{1},\lambda_{2})\delta(x)=-\left\{ \mathcal{L}^{L}(x|\lambda_{1})\overset{\otimes}{,}\kappa^{L}(\lambda_{2})\right\} =-\left\{ g(x)\mathcal{L}^{R}(x|1/\lambda_{1})g^{-1}(x)+J_{1}^{L}(x)\overset{\otimes}{,}g\kappa^{R}(1/\lambda_{2})g^{-1}\right\} =\\
g\otimes g\left(G^{R}(1/\lambda_{1},1/\lambda_{2})-\frac{\lambda_{1}}{1-\lambda_{1}^{2}}\left[C,1\otimes\kappa^{R}(1/\lambda_{2})\right]\right)g^{-1}\otimes g^{-1}\delta(x)\label{eq:GLGR}
\end{multline}
where we used that $\kappa^{R}$ does not depend on the time derivative
of the fields and equation \eqref{eq:commJg}:
\begin{align*}
\left\{ J_{0}^{R}(x)\overset{\otimes}{,}g(y)\right\}  & =\left(1\otimes g\right)C\delta(x-y) & \left\{ J_{0}^{R}(x)\overset{\otimes}{,}g^{-1}(y)\right\}  & =-C\left(1\otimes g^{-1}\right)\delta(x-y)
\end{align*}
Since the $r$-matrices are proportional to $C$ then $g_{1}^{-1}g_{2}^{-1}r_{12}g_{1}g_{2}=r_{12}$.
Using this and \eqref{eq:GLGR} in \eqref{cbYBE-1} we can obtain
the following:
\begin{multline}
\bigl[r_{12}(\lambda_{1},\lambda_{2}),\kappa_{1}^{R}(1/\lambda_{1})\kappa_{2}^{R}(1/\lambda_{2})\bigr]+\\
+\kappa_{1}^{R}(1/\lambda_{1})r_{12}(\lambda_{1},-\lambda_{2})\kappa_{2}^{R}(1/\lambda_{2})-\kappa_{2}^{R}(1/\lambda_{2})r_{12}(\lambda_{1},-\lambda_{2})\kappa_{1}^{R}(1/\lambda_{1})+\\
+\frac{1}{2}\Bigl(G_{12}^{R}(-1/\lambda_{1},1/\lambda_{2})\kappa_{1}^{R}(1/\lambda_{1})-\kappa_{1}^{R}(1/\lambda_{1})G_{12}^{R}(1/\lambda_{1},1/\lambda_{2})-\\
-G_{21}^{R}(-1/\lambda_{2},1/\lambda_{1})\kappa_{2}^{R}(1/\lambda_{2})+\kappa_{2}^{R}(1/\lambda_{2})G_{21}^{R}(1/\lambda_{2},1/\lambda_{1})\Bigr)+\\
+\frac{1}{2}\frac{\lambda_{1}}{1-\lambda_{1}^{2}}\left[\left[C_{12},\kappa_{2}^{R}(1/\lambda_{2})\right],\kappa_{1}^{R}(1/\lambda_{1})\right]_{+}-\frac{1}{2}\frac{\lambda_{2}}{1-\lambda_{2}^{2}}\left[\left[C_{12},\kappa_{1}^{R}(1/\lambda_{1})\right],\kappa_{2}^{R}(1/\lambda_{2})\right]_{+}=0\label{cbYBE-1-1-1}
\end{multline}
Let us see the last two terms
\begin{multline}
\frac{1}{2}\frac{\lambda_{1}}{1-\lambda_{1}^{2}}\left[\left[C_{12},\kappa_{2}^{R}(1/\lambda_{2})\right],\kappa_{1}^{R}(1/\lambda_{1})\right]_{+}-\frac{1}{2}\frac{\lambda_{2}}{1-\lambda_{2}^{2}}\left[\left[C_{12},\kappa_{1}^{R}(1/\lambda_{1})\right],\kappa_{2}^{R}(1/\lambda_{2})\right]_{+}=\\
\frac{1}{2}\left(\frac{\lambda_{1}}{1-\lambda_{1}^{2}}-\frac{\lambda_{2}}{1-\lambda_{2}^{2}}\right)\bigl[C_{12},\kappa_{1}^{R}(1/\lambda_{1})\kappa_{2}^{R}(1/\lambda_{2})\bigr]+\\
+\frac{1}{2}\left(\frac{\lambda_{1}}{1-\lambda_{1}^{2}}+\frac{\lambda_{2}}{1-\lambda_{2}^{2}}\right)\left(\kappa_{1}^{R}(1/\lambda_{1})C_{12}\kappa_{2}^{R}(1/\lambda_{2})-\kappa_{2}^{R}(1/\lambda_{2})C_{12}\kappa_{1}^{R}(1/\lambda_{1})\right)\label{eq:temp1}
\end{multline}
The second line of \eqref{eq:temp1} can be merged with the first
line of \eqref{cbYBE-1-1-1} and the third line of \eqref{eq:temp1}
with the second line of \eqref{cbYBE-1-1-1}:
\begin{align}
r_{12}(\lambda_{1},\lambda_{2})+\frac{1}{2}\left(\frac{\lambda_{1}}{1-\lambda_{1}^{2}}-\frac{\lambda_{2}}{1-\lambda_{2}^{2}}\right)C_{12} & =r_{12}(1/\lambda_{1},1/\lambda_{2})\label{eq:rmatrixInv}\\
r_{12}(\lambda_{1},-\lambda_{2})+\frac{1}{2}\left(\frac{\lambda_{1}}{1-\lambda_{1}^{2}}+\frac{\lambda_{2}}{1-\lambda_{2}^{2}}\right)C_{12} & =r_{12}(1/\lambda_{1},-1/\lambda_{2})\label{eq:rmatrixInvm}
\end{align}
Using this in \eqref{cbYBE-1-1-1} we get
\begin{multline*}
\bigl[r_{12}(1/\lambda_{1},1/\lambda_{2}),\kappa_{1}^{R}(1/\lambda_{1})\kappa_{2}^{R}(1/\lambda_{2})\bigr]+\\
+\kappa_{1}^{R}(1/\lambda_{1})r_{12}(1/\lambda_{1},-1/\lambda_{2})\kappa_{2}^{R}(1/\lambda_{2})-\kappa_{2}^{R}(1/\lambda_{2})r_{12}(1/\lambda_{1},-1/\lambda_{2})\kappa_{1}^{R}(1/\lambda_{1})+\\
+\frac{1}{2}\Bigl(G_{12}^{R}(-1/\lambda_{1},1/\lambda_{2})\kappa_{1}^{R}(1/\lambda_{1})-\kappa_{1}^{R}(1/\lambda_{1})G_{12}^{R}(1/\lambda_{1},1/\lambda_{2})-\\
-G_{21}^{R}(-1/\lambda_{2},1/\lambda_{1})\kappa_{2}^{R}(1/\lambda_{2})+\kappa_{2}^{R}(1/\lambda_{2})G_{21}^{R}(1/\lambda_{1},1/\lambda_{2})\Bigr)=0
\end{multline*}
After changing $1/\lambda_{1}$ and $1/\lambda_{2}$ to $\lambda_{1}$
and $\lambda_{2}$, the last equation is the cbYBE for $\kappa^{R}$.
Therefore we proved that if $\kappa^{L}(\lambda)$ satisfies the cbYBE
then $\kappa^{R}(\lambda)=g^{-1}\kappa^{L}(1/\lambda)g$ also does.

Finally, we prove that equation \eqref{eq:rmatrixInv} follows from
the Poisson algebras of $\mathcal{L}^{L}$ and $\mathcal{L}^{R}$
\eqref{eq:LPalg}:
\begin{align}
\{\mathcal{L}_{1}^{L/R}(x|\lambda_{1}),\mathcal{L}_{2}^{L/R}(y|\lambda_{2})\}=- & \bigl[r_{12}(\lambda_{1},\lambda_{2}),\mathcal{L}_{1}^{L/R}(\lambda_{1})+\mathcal{L}_{2}^{L/R}(\lambda_{2})\bigr]\delta(x-y)+\nonumber \\
+ & \bigl[s_{12}(\lambda_{1},\lambda_{2}),\mathcal{L}_{1}^{L/R}(\lambda_{1})-\mathcal{L}_{2}^{L/R}(\lambda_{2})\bigr]\delta(x-y)-\label{eq:originPoisson}\\
- & 2s_{12}(\lambda_{1},\lambda_{2})\delta'(x-y)\nonumber 
\end{align}
and the inversion property
\[
\mathcal{L}^{L}(\lambda)=g\mathcal{L}^{R}(1/\lambda)g^{-1}+U^{L}
\]
where we used the notation: $U^{L/R}=J_{1}^{L/R}$. Let us start with
the left connections.
\[
\{\mathcal{L}_{1}^{L}(x|\lambda_{1}),\mathcal{L}_{2}^{L}(y|\lambda_{2})\}=\left\{ \left(g\mathcal{L}^{R}(1/\lambda)g^{-1}+U^{L}\right)_{1}(x),\left(g\mathcal{L}^{R}(1/\lambda)g^{-1}+U^{L}\right)_{2}(y)\right\} .
\]
The r.h.s. is equal to the sum of the following three terms
\begin{align}
\Bigl\{\left(g\mathcal{L}^{R}(1/\lambda_{1})g^{-1}\right)_{1}(x), & \left(g\mathcal{L}^{R}(1/\lambda_{2})g^{-1}\right)_{2}(y)\Bigr\}=\nonumber \\
= & g_{1}(x)g_{2}(y)\left\{ \mathcal{L}_{1}^{R}(x|1/\lambda_{1}),\mathcal{L}_{2}^{R}(y|1/\lambda_{2})\right\} g_{1}^{-1}(x)g_{2}^{-1}(y)+\nonumber \\
+ & \frac{\lambda_{1}}{1-\lambda_{1}^{2}}g_{1}g_{2}\left[C_{12},\mathcal{L}_{2}^{R}(1/\lambda_{2})\right]g_{1}^{-1}g_{2}^{-1}\delta(x-y)-\label{eq:tempeq1}\\
- & \frac{\lambda_{2}}{1-\lambda_{2}^{2}}g_{1}g_{2}\left[C_{12},\mathcal{L}_{1}^{R}(1/\lambda_{2})\right]g_{1}^{-1}g_{2}^{-1}\delta(x-y),\nonumber 
\end{align}
\begin{align}
\left\{ \left(g\mathcal{L}^{R}(1/\lambda_{1})g^{-1}\right)_{1}(x),U_{2}^{L}(y)\right\} = & -\frac{\lambda_{1}}{1-\lambda_{1}^{2}}\left(\left[C_{12},U_{1}^{L}\right]\delta(x-y)-C\delta'(x-y)\right),\label{eq:tempeq2}\\
\left\{ U_{1}^{L}(x),\left(g\mathcal{L}^{R}(1/\lambda_{2})g^{-1}\right)_{2}(y)\right\} = & -\frac{\lambda_{2}}{1-\lambda_{2}^{2}}\left(\left[C_{12},U_{1}^{L}\right]\delta(x-y)-C\delta'(x-y)\right).\label{eq:tempeq3}
\end{align}
Let us calculate the first term in the r.h.s of \eqref{eq:tempeq1}.
\begin{align*}
g_{1}(x)g_{2}(y)\left\{ \mathcal{L}_{1}^{R}(x|1/\lambda_{1}),\mathcal{L}_{2}^{R}(y|1/\lambda_{2})\right\}  & g_{1}^{-1}(x)g_{2}^{-1}(y)=\\
- & \bigl[r_{12}(1/\lambda_{1},1/\lambda_{2}),\mathcal{L}_{1}^{L}(\lambda_{1})-U_{1}^{L}+\mathcal{L}_{2}^{L}(\lambda_{2})-U_{2}^{L}\bigr]\delta(x-y)\\
+ & \bigl[s_{12}(1/\lambda_{1},1/\lambda_{2}),\mathcal{L}_{1}^{L}(\lambda_{1})-U_{1}^{L}-\mathcal{L}_{2}^{L}(\lambda_{2})+U_{2}^{L}\bigr]\delta(x-y)\\
- & 2g_{1}(x)g_{2}(y)s_{12}(1/\lambda_{1},1/\lambda_{2})g_{1}^{-1}(x)g_{2}^{-1}(y)\delta'(x-y)
\end{align*}
where we used that $g_{1}g_{2}C_{12}g_{1}^{-1}g_{2}^{-1}=C_{12}$.
The third term can be written as
\begin{multline*}
-2g_{1}(x)g_{2}(y)s_{12}(1/\lambda_{1},1/\lambda_{2})g_{1}^{-1}(x)g_{2}^{-1}(y)\delta'(x-y)=\\
=-2s_{12}(1/\lambda_{1},1/\lambda_{2})\delta'(x-y)+2\left[s_{12}(1/\lambda_{1},1/\lambda_{2}),U_{1}^{L}\right]\delta(x-y)
\end{multline*}
where we used that $\left(f(x)-f(y)\right)\delta'(x-y)=-f'(x)\delta(x-y)$.
Using these the formula above can be written as
\begin{align*}
g_{1}(x)g_{2}(y)\left\{ \mathcal{L}_{1}^{R}(x|1/\lambda_{1}),\mathcal{L}_{2}^{R}(y|1/\lambda_{2})\right\}  & g_{1}^{-1}(x)g_{2}^{-1}(y)=\\
- & \bigl[r_{12}(1/\lambda_{1},1/\lambda_{2}),\mathcal{L}_{1}^{L}(\lambda_{1})+\mathcal{L}_{2}^{L}(\lambda_{2})\bigr]\delta(x-y)+\\
+ & \bigl[s_{12}(1/\lambda_{1},1/\lambda_{2}),\mathcal{L}_{1}^{L}(\lambda_{1})-\mathcal{L}_{2}^{L}(\lambda_{2})\bigr]\delta(x-y)-\\
- & 2s_{12}(1/\lambda_{1},1/\lambda_{2})\delta'(x-y).
\end{align*}
Summing the equations \eqref{eq:tempeq1}, \eqref{eq:tempeq2} and
\eqref{eq:tempeq3}, we can obtain
\begin{align*}
\{\mathcal{L}_{1}^{L}(x|\lambda_{1}),\mathcal{L}_{2}^{L}(y|\lambda_{2})\}=- & \left[\left(r_{12}(1/\lambda_{1},1/\lambda_{2})-\frac{a_{1}-a_{2}}{2}C_{12}\right),\mathcal{L}_{1}^{L}(\lambda_{1})+\mathcal{L}_{2}^{L}(\lambda_{2})\right]\delta(x-y)+\\
+ & \left[\left(s_{12}(1/\lambda_{1},1/\lambda_{2})-\frac{a_{1}+a_{2}}{2}C_{12}\right),\mathcal{L}_{1}^{L}(\lambda_{1})-\mathcal{L}_{2}^{L}(\lambda_{2})\right]\delta(x-y)\\
- & 2\left(s_{12}(1/\lambda_{1},1/\lambda_{2})-\frac{a_{1}+a_{2}}{2}C_{12}\right)\delta'(x-y)
\end{align*}
where we used the following notations
\begin{align*}
a_{1} & =\frac{\lambda_{1}}{1-\lambda_{1}^{2}}, & a_{2} & =\frac{\lambda_{2}}{1-\lambda_{2}^{2}}.
\end{align*}
From the original Poisson bracket \eqref{eq:originPoisson}, we can
see that the $r$- and $s$-matrices satisfies the following identities
\begin{align*}
r_{12}(\lambda_{1},\lambda_{2}) & =r_{12}(1/\lambda_{1},1/\lambda_{2})-\frac{1}{2}\left(\frac{\lambda_{1}}{1-\lambda_{1}^{2}}-\frac{\lambda_{2}}{1-\lambda_{2}^{2}}\right)C_{12},\\
s_{12}(\lambda_{1},\lambda_{2}) & =s_{12}(1/\lambda_{1},1/\lambda_{2})-\frac{1}{2}\left(\frac{\lambda_{1}}{1-\lambda_{1}^{2}}+\frac{\lambda_{2}}{1-\lambda_{2}^{2}}\right)C_{12}.
\end{align*}

\bibliographystyle{elsarticle-num}
\bibliography{on}

\end{document}